\documentclass[preprint]{aastex}
\makeindex

\shorttitle{PLANETARY ACCRETION AROUND M DWARF STARS}
\shortauthors{OGIHARA \& IDA}

\title{\textit{N}-BODY SIMULATIONS OF PLANETARY ACCRETION AROUND M DWARF STARS}
\author{Masahiro Ogihara and Shigeru Ida} 
\affil{Department of Earth and Planetary Sciences, Tokyo Institute of Technology}
\affil{2-12-1 Ookayama, Meguro-ku, Tokyo 152-8551, Japan}
\email{ogihara@geo.titech.ac.jp}

\begin{document}

\begin{abstract}
We have investigated planetary accretion from planetesimals 
in terrestrial planet regions inside the ice line around M dwarf stars 
through \textit{N}-body simulations including tidal interactions with disk gas.
Because of low luminosity of M dwarfs, habitable zones (HZs) are located in 
inner regions ($\sim 0.1$AU).
In the close-in HZ, type-I migration and the orbital decay induced by
eccentricity damping are
efficient according to the high disk gas density 
in the small orbital radii.
Since the orbital decay is terminated around the disk inner edge and
the disk edge is close to the HZ, the protoplanets accumulated near the disk edge
affect formation of planets in the HZ.  
Ice lines are also in relatively inner regions at $\sim 0.3$AU.  
Due to the small orbital radii, icy protoplanets accrete rapidly and 
undergo type-I migration before disk depletion.
The rapid orbital decay, the proximity of the disk inner edge, 
and large amount of inflow of icy protoplanets are characteristic 
in planetary accretion in terrestrial planet regions around M dwarfs.
In the case of full efficiency of type-I migration predicted by the
linear theory, we found that protoplanets that migrate to
the vicinity of the host star undergo close scatterings 
and collisions, and 
4 to 6 planets eventually remain in mutual mean motion 
resonances and their orbits have small
eccentricities ($\la 0.01$) and they
are stable both before and after disk gas decays. 
In the case of slow migration,
the resonant capture is so efficient that densely-packed
$\sim 40$ small protoplanets remain in mutual mean motion resonances.
In this case, they start orbit crossing, after the disk gas decays
and eccentricity damping due to tidal interaction with gas
is no more effective.
Through merging of the protoplanets,
several planets in widely-separated non-resonant orbits with
relatively large eccentricities ($\sim 0.05$) are formed.
Thus, the final orbital configurations (separations,
resonant or non-resonant, eccentricity, distribution)
of the terrestrial planets around M dwarfs sensitively 
depend on strength of type-I migration.
We also found that large amount of water-ice is delivered by type-I migration
from outer regions and final 
planets near the inner disk edge around M dwarfs are generally abundant in water-ice 
except for the innermost one that is shielded by the outer planets, 
unless type-I migration speed is reduced by a 
factor of more than 100 from that predicted by the linear theory.

\end{abstract}
\keywords{methods: N-body simulations - planetary systems - planetary systems: formation - planetary systems: protoplanetary disks -
stars: low-mass, brown dwarfs}

\section{INTRODUCTION}
\label{sec:intro}
Over 300 extrasolar planets have been discovered.
Target stars for exoplanet search were mostly solar-type stars (F, G, K dwarfs), 
although M dwarfs make up 70-80\% of all stars in the galactic disk. 
The low luminosity of M dwarfs 
is disadvantageous for high-dispersion spectroscopic observation, so that radial velocity 
surveys have not discovered large number of planets around M dwarfs. 
However, as improvement of spectroscopic observations, ground-based radial velocity surveys are 
revealing planetary systems around M dwarfs. 
Due to the low luminosity of M dwarfs, habitable zones (HZ),
in which a planet with 
sufficient amount of atmosphere can sustain liquid water 
on its surface, are close to the host stars \citep{kasting_etal93}.
The proximity of the HZs to the host stars 
allow for detection of planets in HZs by current radial velocity observation.
In fact, two planets with minimum masses below $10 M_{\oplus}$ were discovered near the HZ
in a triple planet system around an M star, Gliese 581, with
stellar mass $M_* = 0.31 M_{\odot}$ \citep{udry_etal07}. 
The habitability of these planets (Gl 581c,d) is vigorously under discussion theoretically. 
In addition, gravitational microlensing survey is suited for detection of M dwarf planets, because its 
detection efficiency is independent of stellar luminosity. Most of the planets detected by microlensing 
are orbiting M dwarfs. 
Recent radial velocity and microlensing observations show that Jupiter-mass 
gas giants are generally rare (e.g., \citealt{endl_etal06}, \citealt{johnson_etal07}), 
but Neptune-mass planets are rather abundant (e.g., \citealt{beaulieu_etal06}), 
compared with solar-type stars. 

GJ~436b is the planet 
that was discovered first among Neptune-mass planets
\citep{butler_etal04}.
Transit observations revealed a planet's radius, and its combination with radial velocity measurements 
permits a determination of the planet's density. The evaluated internal density suggests that 
GJ~436b can be composed mainly of ice (\citealt{gillon_etal07}, \citealt{deming_etal07}), 
in spite of proximity to the host star. 
On-going and upcoming transit surveys using space telescopes such as 
Corot, Kepler and TESS, 
besides ground-based transit surveys, 
are expected to reveal lower mass exoplanets around M dwarfs.

``Core accretion'' model (e.g., \citealt{hayashi_etal85}) naturally accounts for
the low abundance of gas giants around M dwarfs, because observationally inferred
low disk mass around M dwarfs inhibits
formation of cores large enough for runaway gas accretion \citep{laughlin_etal04b,ida05}.  
The relatively high abundance of Neptune-mass planets could be accounted for by
truncation of gas accretion at smaller planetary mass due to 
lower disk temperature \citep{ida05}. 
Thus, the Monte Carlo calculation by
\citet{ida05} (at least qualitatively) explained the observed properties in 
the population of gas giants and Neptune-mass planets around M dwarfs.
However, they did not predict abundance of habitable planets,
because it is affected
by type-I migration and detailed orbital configurations 
of close-in planets,
which were not taken into account in their calculation.
\citet{ida08a} included the effect of type-I migration in the similar
calculation, but treatment of close-in planets was still too simple
to discuss the abundance of habitable planets at $\sim 0.1$AU
around M dwarfs.

\textit{N}-body simulation is an efficient tool to address this issue. 
Since physical sizes of planetesimals occupy 
a larger fraction of their Hill radii in the 
terrestrial planet regions ($\sim 0.1~{\rm AU}$) around M dwarfs than around solar-type stars,
strong gravitational scattering is suppressed, which reduces computational cost.
Moreover, we can neglect perturbations from gas giants because
they are rare in the M-dwarf planetary systems, 
which also makes \textit{N}-body simulation simple.

\citet{raymond_etal07} performed a first \textit{N}-body simulation of 
terrestrial planet formation 
from planetary embryos around low-mass stars.
They found that the planets in a HZ may be too small to retain ocean 
because they assumed that disk surface density is proportional to the stellar mass
and the disk model for $1 M_{\odot}$ is the minimum mass
solar nebula (MMSN) model \citep{hayashi81}.
Under this assumption, the isolation mass of the planets is 
proportional to stellar mass (section 2.3).
In HZs, icy grains do not condense in the protoplanetary disk in which gas pressure 
is much smaller than the planetary atmospheric pressure.
One of available sources for the water on the planets is 
delivery of icy planetesimals from the regions beyond 
the ice line \citep{morbidelli_etal00}, although the possibility of forming
H$_2$O through chemical interaction between the planetary magma ocean and
primitive H$_2$ atmosphere is also pointed out \citep{ikoma06}.
Assuming the delivery hypothesis of origin of H$_2$O,  
\citet{raymond_etal07} suggested that the planets in HZs around
M dwarf stars are likely to be dry, 
since radial mixing and therefore water delivery are inefficient in the
lower-mass disks.
\citet{lissauer07} also pointed out the possible lack of
large volatile inventories due to the large collision speeds 
of impacting comets and
substantial mass loss of volatiles due to high activity and 
luminosities of young M dwarfs.

Although the simulations by \citet{raymond_etal07} provided important
insights into probability of habitable planets around M dwarfs, they
neglected the effects of protoplanetary disk gas. 
Since in such inner regions, gas density is so high that 
migration due to gas drag and 
tidal interaction with a gas disk are efficient (see section 2.3)
and accretion timescale of terrestrial planets would be much shorter than 
disk lifetime (see section 2.3), the effects of disk gas play 
important roles in the accretion of planets
in HZs around M dwarfs and water delivery to them.
We will point out in section 3.5 that the planets in
HZs are rather composed mainly of water-ice 
unless type-I migration speed is reduced
by a factor of more than 100 from the linear theory or
the migrating protoplanets are trapped
at an inner boundary of dead zone (\citealt{kretke07}, 
\citealt{ida08b}, \citealt{kretke_etal09}).

With type-I migration,
the proximity of the HZ to the inner disk 
edge would play an important role for final configuration of
planets in HZs, because 
type-I migration stops at the disk edge and planets would accumulate there. 
\citet{terquem07} performed \textit{N}-body simulations 
of protoplanets undergoing type-I migration around solar-type stars. 
Their important finding is that the migrating protoplanets originally formed 
at $\sim 1 {\rm AU}$ interact with the preceding protoplanets near the disk inner edge 
and finally two to five close-in planets remain 
in mutual mean-motion resonances. 
They found that the resonant configuration is maintained
even after disk gas is removed or tidal interaction with the star is added.
They only carried out simulations with type-I migration
with full efficiency that is predicted by the linear theory
(\citealt{ward86}, \citealt{tanaka_etal02}).
We will show that the final orbital configuration is
sensitively dependent of migration speed. 

We thereby carry out \textit{N}-body simulations including the effects of 
damping of orbital eccentricity, inclination and semimajor axis 
(type-I migration) due to disk gas. 
Although \citet{raymond_etal07} and \citet{terquem07} start their
simulations from planetary embryos that have already grown to their
isolation masses, our simulation starts from many small planetesimals,
taking fully into account their gravitational interactions.
Our calculations also cover a broad range of orbital radius 
from the disk inner edge to beyond the ice line. 
Since there is still uncertainty in the type-I migration speed, 
we perform both simulations with and without type-I migration.

In section~\ref{sec:model}, we describe the disk model, the formulas of forces for 
aerodynamic and gravitational gas drag, and calculation methods.
The results of \textit{N}-body simulations of planetary accretion are shown in section~\ref{sec:results}.
In section~\ref{subsec:comp}, we discuss water delivery to inner planets around M dwarfs.
Section~\ref{sec:conc} is devoted to the conclusion and discussion.

\section{MODEL AND CALCULATION METHODS}
\label{sec:model}

Here, we consider an M5V type star with mass $M_* = 0.2 M_\odot$ and luminosity 
$L_* \simeq 0.01 L_{\odot}$, using a mass-luminosity 
relation ($L_* \propto M_*^3$). 
This relation roughly fits observational data 
for a range of 0.1 to 1 $M_{\odot}$ stars in main-sequence stages
(e.g., \citealt{habets81}, \citealt{scalo_etal07}).
Although pre-main-sequence stages are relatively long for 
the low mass stars and luminosity is relatively high during the
pre-main-sequence stages,
we will use HZs determined by the main-sequence radiation,
because our main purpose is to clarify the dynamics and
accretion process among protoplanets that have migrated to
the regions near disk inner edges.
The simulation with evolving HZs and ice lines 
due to the luminosity evolution is left to a future work. 

We integrate the orbits of planetesimals, taking into account
their merging by direct collisions,
their gravitational interactions, aerodynamic gas drag, 
``gravitational drag'' (damping of orbital eccentricity and inclination
due to tidal interaction with a gas disk), and type-I migration. 
The models for surface densities of disk gas and an 
initial planetesimal swarm are explained in section 2.1.
Basic equations for orbital integration and initial conditions
are presented in section 2.2.
Although the detailed expressions for the aerodynamic gas and
gravitational drag forces are given in Appendixes A and B,
we summarize their characteristic timescales in section 2.3, as well as
the timescale of planetesimal accretion, which are
useful to understand the results of the \textit{N}-body simulations.

\subsection{Disk Model}

Following \citet{ida04},
we scale the gas surface density $\Sigma_g$ of disks as
\begin{eqnarray}
\Sigma_g = 2400 f_g\Bigl(\frac{r}{1\mathrm{AU}}\Bigr)^{-3/2} \,\mathrm{g\, cm}^{-2},
\label{eq:Sigma}
\end{eqnarray}
where $f_g$ is a scaling factor; $\Sigma_g$ is 1.4 times of those in the MMSN model if $f_g = 1$.
Because current observations cannot strictly constrain the radial
gradient of $\Sigma_g$, we here assume $f_g$ is constant with $r$ except for the inner edge.
For disks around stars with $M_* \sim 1 M_{\odot}$, 
the observationally inferred averaged value of $f_g$ is 
$\sim 1$ although the values
have dispersion of two orders of magnitude (see discussion in \citealt{ida04}).
We here adopt $f_g = 1$ for the disks around the $M_* = 0.2 M_\odot$ star. 
Although averaged $f_g$ may be several times smaller for these stars,
$f_g = 1$ is within the two orders of magnitude dispersion.
The relatively large value of $f_g$ is to study possible 
formation of habitable planets, which are large enough to 
retain water on their surface, around low mass stars.
In section 3.5, we will discuss how the results are affected 
if less massive disks ($f_g \sim 0.2$), which may be
averaged disks around these stars, are considered.

We assume the temperature distribution of 
an optically thin disk \citep{hayashi81},
\begin{eqnarray}
T = 2.8 \times 10^2 f_g \Bigl(\frac{r}{1\mathrm{AU}}\Bigr)^{-1/2} \Bigl(\frac{L_*}{L_\odot}\Bigr)^{1/4} \, \mathrm{K}.
\label{eq:temp}
\end{eqnarray}
Corresponding sound velocity $c_s$ and disk scale height $h$ are 
\begin{eqnarray}
c_s = 1.0 \times 10^5 \Bigl(\frac{r}{1\mathrm{AU}}\Bigr)^{-1/4} \Bigl(\frac{L_*}{L_\odot}\Bigr)^{1/8} \, \mathrm{cm \, s^{-1}},
\label{eq:sound_vel}
\end{eqnarray}
\begin{eqnarray}
h \simeq 4.7 \times 10^{-2} \Bigl(\frac{r}{1\mathrm{AU}}\Bigr)^{5/4} \Bigl(\frac{L_*}{L_\odot}\Bigr)^{1/8} 
\Bigl(\frac{M_*}{M_\odot}\Bigr)^{-1/2} \, \mathrm{AU}.
\label{eq:scale_hight}
\end{eqnarray}

In the current paradigm, T Tauri disks are truncated by the stellar magnetosphere within the corotation radius, 
with materials accreting along magnetic field lines onto high-latitude regions of the star. 
The corotation radius $r_{\rm corot}$ is the radius at which the Keplerian orbital period in the disk equals the stellar rotation period ($P$),
\begin{eqnarray}
r_{\rm corot} = 0.04 \Bigl(\frac{P}{3\mathrm{days}}\Bigr)^{2/3} \Bigl(\frac{M_*}{M_\odot}\Bigr)^{1/3} \, \mathrm{AU}.
\end{eqnarray}
Here, we set the location of disk inner edge at 0.05~AU.
From Eq.~(\ref{eq:scale_hight}), the scale height of a disk at 0.05~AU around a star with $M_* = 0.2 M_{\odot}$
is $\sim 1.4 \times 10^{-3}$~AU.
We assume that the gas surface density of the disk (equivalently $f_g$) smoothly vanishes at 0.05~AU with 
a hyperbolic tangent function with the width of $10^{-3}$~AU, which
is comparable to $h$.
When planets enter the inner cavity, they do not feel gas drag 
any more and their inward migration ceases because the drag and migration
rates are proportional to disk gas surface density.
In some runs, we reverse the direction of the migration near the edge 
according to positive pressure gradient there \citep{tanaka_etal02,masset_etal06}.

According to gas surface density given by Eq.~(\ref{eq:Sigma}),
we scale the surface density of a planetesimal disk with a
scaling factor, $f_d$, as
\begin{eqnarray}
\Sigma_d = 10 \eta_{\rm ice} f_d \Bigl(\frac{r}{1\mathrm{AU}}\Bigr)^{-3/2} \,\mathrm{g\, cm}^{-2}.
\label{eq:solid}
\end{eqnarray}
For solar metallicity, $f_d = f_g$, so we use $f_d =1$.
The factor $\eta_{\rm ice}$ expresses the increase of solid materials 
due to ice condensation outside the ``ice line.'' 
The location of the ice line is determined 
in such a way that the disk temperature equals the ice condensation 
temperature.  Assuming the condensation temperature of 170~K, 
the location of the ice line is derived from 
Eq.~(\ref{eq:temp}):
\begin{eqnarray}
r_{\rm ice} \simeq 2.7 \Bigl(\frac{L_*}{L_{\odot}}\Bigr)^{1/2} \, \mathrm{AU}.
\end{eqnarray}
Since we consider stars with $L_* \sim 0.01M_{\odot}$,
we set $r_{\rm ice} = 0.3 \mathrm{AU}$.
In the MMSN model, $\eta_{\rm ice} = 4.2$ at $r > r_{\rm ice}$. \citet{pollack_etal94} derived $\eta_{\rm ice} \simeq 3$.
In the standard set of our simulations, we adopt $\eta_{\rm ice}$ as
\begin{eqnarray}
\eta_{\rm ice} = \Bigl \{
\begin{array}{ll}
1 & [r < 0.3 {\rm AU}]\\
3 & [0.3 {\rm AU} < r ].
\end{array}
\Bigr. 
\end{eqnarray}
But, in some runs, we adopt a higher value for $r_{\rm ice}$ (see section 3.3).

\subsection{Orbital Integration and Initial Conditions} 

We integrate the orbits of planetesimals
with 4th-order Hermite scheme \citep{makino92} and 
hierarchical individual timestep \citep{makino91}. 
The basic equations of motions of particle $k$ at $\textbf{\textit{r}}_k$ 
in heliocentric coordinates are
\begin{eqnarray}
\frac{d^2 \textbf{\textit{r}}_k}{dt^2} 
& = & -GM_* \frac{\textbf{\textit{r}}_k}{ |\textbf{\textit{r}}_k|^3} 
- \sum_{j \neq k} GM_j 
\frac{\textbf{\textit{r}}_k - \textbf{\textit{r}}_j}{|\textbf{\textit{r}}_k - \textbf{\textit{r}}_j|^3} 
\nonumber\\
  &   &   
+ \textbf{\textit{F}}_{\rm aero} + \textbf{\textit{F}}_{\rm damp} + \textbf{\textit{F}}_{\rm mig},
\label{eq:eq_motion} 
\end{eqnarray}
where $k,j = 1,2,...$, the first term is gravitational force of the central star and
the second term is mutual gravity between the bodies. 
We calculate the self gravity directly summing up interactions of all pairs on the special-purpose 
computer for \textit{N}-body simulation, GRAPE-6.
$\textbf{\textit{F}}_{\rm aero}$, $\textbf{\textit{F}}_{\rm damp}$ and $\textbf{\textit{F}}_{\rm mig}$ are 
specific forces due to aerodynamic gas drag, gravitational drag, and
type-I migration, the detailed expressions of which are described in 
Appendixes~\ref{ap:aero} and \ref{ap:grav}. 
We neglect the indirect term since the total mass of the planetesimals is $\sim 10^{-4}$ times the mass of the central star.

When physical radii of two bodies overlap, perfect accretion is assumed. After the collision, a new body is created, 
conserving total mass and momentum of the two colliding bodies. 
The physical radius of a body is 
determined by its mass $M$ and internal density $\rho_{\rm p}$ as
\begin{eqnarray}
r_{\rm p} = \Bigl(\frac{3}{4 \pi} \frac{M}{\rho_{\rm p}}\Bigr)^{1/3}.
\end{eqnarray}
We adopt a realistic value $3~{\rm gcm}^{-3}$ for $\rho_{\rm p}$.

Initially 5,000 planetesimals are placed between 0.05~AU (disk inner edge) and 0.4~AU. 
To study accretion process of terrestrial 
planets inside the ice line in more detail, we initially set
more bodies inside the ice line (3898 bodies
with mass $2.3 \times 10^{24}~{\rm g}$) than outside it 
(1102 bodies with mass $6.5 \times 10^{24}~{\rm g}$) 
in the nominal simulations. 
The initial velocity dispersion of the bodies is set to be their escape velocity.
The corresponding initial eccentricity 
and inclination are given by $v_{\rm esc} = \sqrt{e^2 + i^2} v_{\rm K}$
with $e = 2i$ \citep{ida92}.
Table~\ref{tbl:initial} shows simulation conditions for individual runs and final results.
In the first set (setA) of runs (runA1 - runA4), we include all the damping of $e$, $i$ and $a$ due to aerodynamic and 
gravitational drag and type-I migration, 
while type-I migration is neglected in the second set (setB: runB1 - runB4).
Because coagulation to final planets may be a stochastic process, 
for each set we performed 4 runs with different 
seeds for random numbers to generate initial 
angular distribution of planetesimals.

In addition to these sets, we also carried out runs with reversed torque of type-I migration near the disk
inner edge (setC: runC1 - runC2) and runs with
$\eta_{\rm ice} = 14$ at $r > r_{\rm ice}$ (setD: runD1 - runD4 
with type-I migration, setE: runE1 - runE4 without type-I migration), 
which are described in section~3.3.

\subsection{Characteristic Timescales}

To understand the results of \textit{N}-body simulations,
we here summarize the timescales of gas and gravitational
drag, type-I migration and planetesimal accretion.
The characteristic damping timescale for orbital eccentricity ($e$) 
and inclination ($i$) of an isolated planetesimal
due to aerodynamic gas drag is (for more detailed
expressions, see \citealt{adachi_etal76})
\begin{eqnarray}
t_{\rm aero} \sim \frac{\Delta u}{F_{\rm aero}} 
  & \simeq 3.4 \times 10^6 
   \left( \frac{((5/8) e^2 + (1/2) i^2 + \eta ^2 )^{1/2}}{0.01} \right)^{-1} 
\Bigl(\frac{M}{M_{\oplus}}\Bigr)^{1/3} 
\Bigl(\frac{M_*}{0.2 M_{\odot}}\Bigr)^{-1/2} \nonumber\\
  & \times \Bigl(\frac{\rho_{\rm p}}{3\mathrm{g\, cm}^{-3}}\Bigr)^{2/3} 
    \Bigl(\frac{a}{1\mathrm{AU}}\Bigr)^{13/4} \, \mathrm{years},
\label{eq:time_aero}
\end{eqnarray}
where $F_{\rm aero}$ is specific drag force acting on the planetesimal
(eq.~[\ref{eq:f_aero}]) and $a$, $M$ and 
$\rho_{\rm p}$ are its semimajor axis, mass 
and bulk density, respectively.
The relative velocity between the planetesimals
and disk gas, $\Delta u$, is given by
$\simeq ((5/8) e^2 + (1/2) i^2 + \eta ^2)^{1/2} v_{\rm K}$, 
where $v_{\rm K}$ is Keplerian velocity.
The velocity of disk gas $v_{\rm gas}$ 
is smaller than 
Kepler velocity $v_{\rm K}$ by a fraction (eq.~[\ref{eq:eta}]),
\begin{equation}
\eta \simeq \frac{v_{\rm K}-v_{\rm gas}}{v_{\rm K}} = 2.8 \times 10^{-3} \Bigl(\frac{r}{1\mathrm{AU}}\Bigr)^{1/2}
\Bigl(\frac{M_*}{0.2 M_{\odot}}\Bigr)^{-1} 
\Bigl(\frac{L_*}{0.01 L_{\odot}}\Bigr)^{1/4}. 
\end{equation}
The timescale for damping of semimajor axis ($a$) is
\begin{eqnarray}
t_{{\rm aero},a} = \frac{a}{\dot{a}} \sim \frac{t_{\rm aero}}{2 \eta} = & 0.6 \times 10^9 \Bigl(\frac{C_{\rm D}}{0.5}\Bigr)^{-1} f_g^{-1}
\Bigl(\frac{((5/8) e^2 + (1/2) i^2 + \eta ^2)^{1/2}}{0.01}\Bigr)^{-1} \Bigl(\frac{M}{M_{\oplus}}\Bigr)^{1/3} 
\nonumber \\
 & \times 
\Bigl(\frac{\rho_{\rm p}}{3\mathrm{g\, cm}^{-3}}\Bigr)^{2/3} \Bigl(\frac{a}{1\mathrm{AU}}\Bigr)^{11/4}
\Bigl(\frac{M_*}{0.2 M_{\odot}}\Bigr)^{1/2} 
\Bigl(\frac{L_*}{0.01 L_{\odot}}\Bigr)^{-1/4} 
\, \mathrm{years}.
\label{eq:time_mig_aero}
\end{eqnarray}
At $a \sim 0.1$AU, $t_{{\rm aero},a} \sim 1.5 \times 10^6 f_g^{-1}$ years
even for a Mars-mass planet ($M \sim 0.1M_{\oplus}$), so that gas drag
cannot be neglected even for protoplanets.
For small planetesimals, the gas drag is more important.

A planet gravitationally perturbs the disk gas and excites density waves. 
The waves damp the eccentricity, the inclination 
(e.g., \citealt{ward93}, \citealt{artymowicz93}) and the semimajor axis of the planet (e.g., \citealt{goldreich80}, \citealt{ward86}). 
The detailed expressions of gravitational gas drag forces $\textbf{\textit{F}}_{\rm damp}$ and $\textbf{\textit{F}}_{\rm mig}$
are given in Appendix~\ref{ap:grav}.
Orbital eccentricities and inclinations are damped by both torques from inner and outer disks in a similar way 
to dynamical friction from planetesimals. The damping timescales are \citep{tanaka04}
\begin{eqnarray}
t_{\rm damp} = - \frac{e}{\dot{e}} 
             & \sim & \Bigl(\frac{M}{M_*}\Bigr)^{-1} 
               \Bigl(\frac{\Sigma_g a^2}{M_*}\Bigr)^{-1}
               \Bigl(\frac{c_s}{v_{\rm K}}\Bigr)^4 \Omega_{\rm K}^{-1} \\
             & \simeq & 70 f_g^{-1} \Bigl(\frac{M}{M_{\oplus}}\Bigr)^{-1} 
                \Bigl(\frac{a}{1{\rm AU}}\Bigr)^2 
                \Bigl(\frac{M_*}{0.2 M_{\odot}}\Bigr)^{-1/2}
                \Bigl(\frac{L_*}{0.01 L_{\odot}}\Bigr)^{1/2} \,{\rm years}.
\label{eq:t_damp}
\end{eqnarray}
On the other hand, secular inward migration due to tidal interaction with disk gas, that is known as 
``type-I migration" is caused by torque imbalance. 
The migration timescale is 
\begin{eqnarray}
t_{\rm mig} = - \frac{a}{\dot{a}} 
& \simeq & \frac{1}{2.7 + 1.1 q}
\Bigl(\frac{M}{M_*}\Bigr)^{-1}\Bigl(\frac{\Sigma_g a^2}{M_*}\Bigr)^{-1}\Bigl(\frac{c_s}{v_K}\Bigr)^2 \Omega^{-1} 
\label{eq:t_mig_2}\\ 
& = & 7.1 \times 10^3 f_g^{-1}
\Bigl(\frac{M}{M_{\oplus}}\Bigr)^{-1} \Bigl(\frac{a}{1 \, \mathrm{AU}}\Bigr)^{3/2} \Bigl(\frac{M_*}{0.2 M_{\odot}}\Bigr)^{1/2}
\Bigl(\frac{L_*}{0.01 L_{\odot}}\Bigr)^{1/4} \, \mathrm{years}, 
\label{eq:t_mig}
\end{eqnarray}
assuming the gas surface density is proportional to $a^{-q}$. Here we used the MMSN density profile, 
$\Sigma_g \propto a^{-1.5}$, to derive Eq.~(\ref{eq:t_mig}).
In runC1 and runC2, the effect of negative $q$ at the inner edge, 
which reverses the tidal torque, is taken into account.

Since type-I migration is resulted in by torque imbalance, 
non-linear effects could change migration rate significantly.
Therefore we also perform simulations in which 
bodies are free of type-I migration (runB1 - runB4).
Note that even in this set, the migration induced by 
damping of $e$ due to the gravitational drag
exists (section 3.1.2) and the runs in this set are comparable to
the runs with type-I migration of $\sim 100$ times reduced speed.

Growth rate of a protoplanet with mass $M$ and physical radius
$r_{\rm p}$ due to accretion of small planetesimals
is estimated simply by
\begin{eqnarray}
\frac{dM}{dt} \sim \Sigma_d \pi r_{\rm p}^2 \Bigl(1 + \frac{v^2_{\rm esc}}{v^2_{\rm ran}}\Bigr) \Omega_{\rm K},
\end{eqnarray}
where $v_{\rm ran} \simeq (e^2 + i^2)^{1/2} v_{\rm K}$.
With more detailed formula and 
$v_{\rm ran}$ that is determined by balance between scattering of
the planetesimals by the protoplanet and gas drag to them,
the accretion timescale is expressed as \citep{ida04}
\begin{eqnarray}
\label{eq:t_acc}
t_{\rm acc} \simeq 1.2 \times 10^6 \eta_{\rm ice}^{-1} 
f_d^{-1} f_g^{-2/5} 
\Bigl(\frac{a}{1 {\rm AU}}\Bigr)^{27/10} \Bigl(\frac{M}{M_{\oplus}}\Bigr)^{1/3}
\Bigl(\frac{M_*}{0.2 M_{\odot}}\Bigr)^{-1/6} 
\Bigl(\frac{m}{10^{24} {\rm g}}\Bigr)^{2/15} 
\Bigl(\frac{\rho_{\rm p}}{3\mathrm{g\, cm}^{-3}}\Bigr)^{1/3}
\,{\rm years}.
\end{eqnarray}

The growth timescale of terrestrial planets around M dwarfs is shorter 
than that around solar-type stars because of small $a$.
Hence, it is reasonable to assume that full amount of disk gas exists during the terrestrial planet formation
because depletion timescale of disks around M dwarfs may be $\sim 10^6-10^7$ years or more. 
On the other hand, the damping timescales are short in inner regions 
where the gas density is high.
As stated in section 1, 
the effects of disk gas cannot be neglected
when we consider accretion of planets
in HZs around M dwarfs.

The isolation mass, which is the mass of all 
the solid materials in a feeding zone of the protoplanets, is 
\begin{eqnarray}
\label{eq:m_iso}
M_{\rm iso} \simeq 0.23 \eta_{\rm ice}^{3/2} f_d^{3/2} 
\Bigl(\frac{a}{1 \mathrm{AU}}\Bigr)^{3/4} 
\Bigl(\frac{\Delta a}{7.5 r_{\rm H}}\Bigr)^{3/2} 
\Bigl(\frac{M_*}{0.2M_{\odot}}\Bigr)^{-1/2} M_{\oplus},
\end{eqnarray}
where $\Delta a$ is width of the feeding zone.
The mutual Hill radius 
$r_{\rm H}$ is defined by
\begin{equation}
\label{eq:hill}
r_{\rm H} = \Bigl(\frac{M_1 + M_2}{3 M_*}\Bigr)^{1/3} \frac{M_1 a_1 + M_2 a_2}{M_1 + M_2}.
\end{equation}
where $M_1$ and $M_2$ are masses of interacting bodies
that we are concerned with and $a_1$ and $a_2$ are their semimajor axes.  
In Eq.~({\ref{eq:m_iso}), $M_1 = M_2 = M_{\rm iso}$ is assumed.  
The value of $\Delta a = 7.5 r_{\rm H}$ is a typical value
obtained by \textit{N}-body simulation at $a \sim 0.1$AU, which is
slightly smaller than the value obtained at $\sim 1$ AU
\citep{kokubo98,kokubo02}.
The isolation mass is the maximum mass that the planet can
acquire through ``runaway/oligarchic'' growth
(accretion of planetesimals) before onset of orbit crossing
and coagulation among the isolated protoplanets.

The migration timescale for the protoplanet 
with mass $M_{\rm iso}$ is 
\begin{equation}
t_{\rm mig,iso} = 3.1 \times 10^4 
\eta_{\rm ice}^{-3/2} f_d^{-3/2} f_g^{-1}
\Bigl(\frac{a}{1 \, \mathrm{AU}}\Bigr)^{3/4} 
\Bigl(\frac{L_*}{0.01 L_{\odot}}\Bigr)^{1/4} \, \mathrm{years}. 
\label{eq:t_mig_iso}
\end{equation}
Its accretion timescale is
\begin{equation}
t_{\rm acc,iso} = 0.70 \times 10^6 
\eta_{\rm ice}^{-1/2} f_d^{-1/2} f_g^{-2/5}
\Bigl(\frac{a}{1 \, \mathrm{AU}}\Bigr)^{59/20} 
\Bigl(\frac{M_*}{0.2 M_{\odot}}\Bigr)^{-1/3}
\Bigl(\frac{\rho_{\rm p}}{3\mathrm{g\, cm}^{-3}}\Bigr)^{1/3}
\, \mathrm{years}.
\label{eq:t_acc_iso}
\end{equation}
At $a \la 0.3$AU that we are concerned with in this paper,
both $t_{\rm mig,iso}$ and $t_{\rm acc,iso}$ are significantly
shorter than disk lifetime of $\sim 10^6-10^7$ years for
$f_g \sim f_d \sim 1$.
This is also the case even for small-mass disks with
the surface density 5 times smaller than that of the MMSN
($f_d = f_g \simeq 0.15$), which 
\citet{raymond_etal07} considered.
Thus, type-I migration cannot be neglected even for 
the small-mass disks, unless non-linear effects
slow down or halt the migration significantly (section 3.5).

\section{RESULTS}
\label{sec:results}
\subsection{Runaway/oligarchic Growth before Disk Gas Depletion}
\subsubsection{Case with Type-I Migration}
\label{subsec:type-I}

The result for a typical run including the effect of type-I migration
(runA1) is shown in Figure~\ref{fig:snap_run1}.
In the snapshots, the sizes of circles 
are proportional to the physical radii of the bodies, and
$\textit{T}_{\rm K}$ is the Kepler time at 0.1~AU around 
a $0.2 M_{\odot}$ star, which is $\simeq 0.071$ year. 
Since the accretion timescale depends strongly on semimajor axis $a$
(Eq.~[\ref{eq:t_acc}]) and it is shorter for smaller $a$,
accumulation of planetesimals proceeds in an inside-out manner.
Equation~(\ref{eq:t_acc}) for $M \sim M_{\rm iso} \sim 0.04 M_{\oplus}$
at 0.1~AU is $10^3~{\rm yrs} = 1.4 \times 10^4~T_{\rm K}$, 
which agrees with the result in Fig.~\ref{fig:snap_run1}.

According to growth of a protoplanet,
the type-I migration timescale becomes shorter, while 
the accretion timescale becomes longer.
Thereby, the protoplanet starts migration when
its mass exceeds the critical mass beyond which $t_{\rm mig} < t_{\rm acc}$.
From Eqs.~(\ref{eq:t_mig}) and (\ref{eq:t_acc}), the critical mass is
\begin{eqnarray}
M_{\rm crit, mig} \simeq 4.0 \times 10^{-2} \eta_{\rm ice}^{3/4} f_d^{3/4} f_g^{-9/20} 
\Bigl(\frac{a}{1 \mathrm{AU}}\Bigr)^{-9/10} 
\Bigl(\frac{M_*}{0.2 M_{\odot}}\Bigr)^{1/2}
\Bigl(\frac{L_*}{0.01 L_{\odot}}\Bigr)^{3/16}
M_{\oplus}. 
\label{eq:M_crit_mig}
\end{eqnarray}
Because at $\sim 0.1$AU this value is close to the isolation mass,
protoplanets start migration after they accrete most of planetesimals
in their feeding zone.
On the other hand, in outer regions, since $M_{\rm crit, mig} < M_{\rm iso}$,
protoplanets start migration leaving behind large amount of planetesimals.
Most of the initial mass is finally concentrated near the disk inner edge 
after $10^6 \textit{T}_{\rm K}$, leaving few planetesimals in outer regions.
Although eccentricities of the bodies are excited by close encounters 
with neighbor bodies during the growth stage, 
final eccentricities of the planets are kept small ($\la 0.01$) 
due to gravitational drag.
Resonant effect can raise the eccentricities, however, gas drag 
damps them significantly because of the high gas surface density.

Orbital evolution of runA1 is shown in Fig.~\ref{fig:t_a_run1}. In this figure, 
the most massive 30 planets at each time are plotted as circles
(there are many other small bodies). 
Accretion takes place more rapidly in inner region than in outer region at 
the beginning of the simulation.
The protoplanets with $M > M_{\rm crit, mig}$ 
undergo inward type-I migration.
Since $t_{\rm mig}$ is shorter at smaller orbital radius,
the migration is accelerated until the protoplanets reach the disk inner edge.
The firstly migrated protoplanets interact with each other.
After some merging, the survived protoplanets are
captured in mutual mean motion resonances. 
After that, successively migrated planets interact with
outermost planets that have accumulated near the disk edge.
The interaction mostly results in merging of the planets and
the merger is again captured in a mean motion resonance. 
Note that the number of remaining planets near the edge is almost 
constant (4-6 planets) during this phase, 
although protoplanets that formed in outer regions migrate to interact with
the close-in planets one after another.
In this run, after $\sim 5 \times 10^6 \textit{T}_{\rm K}$, 
any more protoplanets which are large enough to affect the inner planets 
do not approach to the inner planetary system, so that this run 
ends up with a stable configuration consisting of six planets with $M > 0.01 M_{\oplus}$
near the disk inner edge. 
Most of the final planets are pushed into the cavity at $a < 0.05$AU,
in which type-I migration
is no more effective, by the outer planets that keep loosing angular 
momentum by type-I migration.
The largest planet in the final state is the third innermost planet, 
the mass of which is $0.63 M_{\oplus}$. 
Note that this value is more than 10 times of the values of $M_{\rm crit,mig}$
($\sim 0.01-0.06M_{\oplus}$ at 0.05-0.4AU).
This implies that coagulation near the disk edge is so efficient.
The value of $0.63 M_{\oplus}$ is also more than 25 times larger than
$M_{\rm iso}$ ($\sim 0.024 M_{\oplus}$) at 0.05AU. 
Without migration, such large planets cannot accrete at $\sim 0.05$AU.

All the final six planets are trapped in first-order commensurability 
(mean-motion resonances) with the planets which lie next to the planets. 
For example, the innermost pair (a pair of the innermost and the second 
innermost planets) has 6:5 commensurability, and the second pair 
(a pair of the second and the third innermost planets) 
has 5:4 commensurability.
The perturbations during one passage rapidly increase for
$\Delta a \la 5 r_{\rm H}$ \citep{ida90}.
Since resonant trapping at $\Delta a > 5 r_{\rm H}$ may not be
resistant against perturbations from other bodies other than the pair,
it is expected that the pair may be eventually trapped at
mean-motion resonances close to $\Delta a \sim 5r_{\rm H}$.
The obtained orbital separations of the $i$-th pairs ($i = 1,2,..,5$)
are $\Delta a \simeq 7.3 r_{\rm H}, 8.3 r_{\rm H}, 6.8 r_{\rm H}, 5.4 r_{\rm H}$ 
and $8.7 r_{\rm H}$, respectively. 
As will be shown in section 3.2, these final orbital configuration
is stable even if disk gas is removed.

All the runs in  setA with the effect of type-I migration 
end in very similar results: 
Most of the protoplanets (large planetesimals) pile up near the inner disk edge, 
and in final state, most of the planets are captured 
in mean motion resonances after coagulation and close scattering among the
protoplanets.
The average number of the final planets in
runA1 - runA4 is $\overline N = 5.0 \pm0.71$, 
which is comparable to that obtained by \citet{terquem07}.
The average mass of the largest planet is 
$\overline{M}_{\rm max} = 0.82 \pm 0.17 M_{\oplus}$, which is significantly
larger than $M_{\rm iso}$ at 0.05AU and
$M_{\rm crit,mig}$ at 0.05-0.4AU.

\subsubsection{Case without Type-I Migration}
\label{subsec:without_type-I}
A typical result excluding the effect of type-I migration in setB is shown in
Figure~\ref{fig:snap_model59}. 
The figure shows snapshots of the system 
on the $a-e$ planes for runB1.
Although the effect of type-I migration is not included, 
planets tend to migrate inward. 
The inward migration is induced by eccentricity damping 
by tidal interaction with disk gas.
Since the angular momentum 
$(L = \sqrt{G M_* a (1-e^2)})$ is almost conserved, 
damping of eccentricity yields damping of 
semimajor axis. 
The damping timescale of semimajor axis is roughly given by 
\begin{eqnarray}
\label{eq:damp_a}
t_{{\rm damp},a}  &=&  - \frac{a}{\dot{a}} = \frac{1}{2 e \dot{e}}  = \frac{t_{\rm damp}}{2 e^2}\\
\label{eq:damp_a_2}
 & \simeq&  0.7 \times 10^6 f_g^{-1} \Bigl(\frac{e}{0.01}\Bigr)^{-2} \Bigl(\frac{M}{M_{\oplus}}\Bigr)^{-1} 
\Bigl(\frac{a}{1 {\rm AU}}\Bigr)^2 
\Bigl(\frac{M_*}{0.2 M_{\odot}}\Bigr)^{-1/2}
\Bigl(\frac{L_*}{0.01 L_{\odot}}\Bigr)^{1/2}
\, {\rm years}, 
\end{eqnarray}
where Eq.~(\ref{eq:t_damp}) was used.
The values of $e$ are typically $\sim 0.01$ in our simulations,
but they change with time and depend on locations.
Compared to Eq.~(\ref{eq:time_mig_aero}), 
we find that the semimajor axis damping induced from the eccentricity damping 
is more effective than aerodynamic gas drag for
$M \sim 0.01-1M_{\oplus}$, even if 
the uncertainty in the values of $e$ is taken into account.
Since this timescale is about 100 times longer than $t_{\rm mig}$
(Eq.~[\ref{eq:t_mig}]) at $a \sim 0.1$AU, this calculation is equivalent to
the case with 100 times reduced type-I migration speed.
Nevertheless, this migration timescale is of the order 
of $10^5 {\rm yrs}$ for a planet of 
$0.1 M_{\oplus}$ at 0.1~AU around a star of $0.01 L_{\odot}$, 
which is considerably shorter than disk lifetime.
This effect plays an important role 
in planetary formation in HZs around M dwarfs,
although it can be neglected in HZs ($a \sim 1$AU) around G dwarfs.
This migration phenomenon can be seen in previous works including tidal $e$-damping with gas disk \citep{ogihara_etal07}.
In Fig.~\ref{fig:snap_model59}, the planets have a tendency to move inward. However, several planets remain in
outside regions whereas in runA1 few planets remained outside region.
The final eccentricities are damped down to 0.01 as in the previous set.

Figure~\ref{fig:t_a_model59} is the same plots as Fig.~\ref{fig:t_a_run1} 
for runB1.
In this run, the resonant capture is so effective because of the slow 
migration that about 45 planets with mass larger than $0.01~M_{\oplus}$
are lined up in resonances with little coagulation among them.
On the other hand, in runA1, migrated planets were not readily trapped at a first
encounter and settled to resonant configurations after 
close scattering and coagulation.
In runB1, the largest planet is located at 0.21~AU, the mass of which is $0.21~M_{\oplus}$. 
Almost all the planets are captured in mean motion resonances, however, 
they tend to have closer commensurabilities, such as 14:13, than those for runA1.
In the case with type-I migration (setA), successively migrated planets interact with 
outermost planets, leading to merger of planets. 
In this case, however, perturbations from many other migrated planets tend to 
break resonant capture at distant separation, 
resulting in capture in closer (stronger) 
mean motion resonances.

All the simulations in this set (runB1 - runB4) exhibit qualitatively the same results.
The average number of final planets is 
as large as $\overline N = 40 \pm 3.3$, 
which is about one order larger than that in setA. 
The average mass of the largest planet is $\overline M_{\rm max} = 0.20 \pm0.033 M_{\oplus}$.

The critical planet mass beyond which $t_{{\rm damp},a} < t_{\rm acc}$ is 
\begin{eqnarray}
\label{eq:m_crit}
M_{\rm crit,damp} \simeq \eta_{\rm ice}^{3/4} f_d^{3/4} f_g^{-9/20} 
\Bigl(\frac{e}{0.01} \Bigr)^{-3/2} \Bigl(\frac{a}{1 \mathrm{AU}}\Bigr)^{-21/40} 
\Bigl(\frac{M_*}{0.2 M_{\odot}}\Bigr)^{-1/4} \Bigl(\frac{L_*}{0.01 L_{\odot}}\Bigr)^{3/8}
M_{\oplus}. 
\end{eqnarray}
This value is greater than the isolation mass (Eq.~[\ref{eq:m_iso}])
at $a \la 2$AU, so that all the protoplanets in the simulations in this set grow up 
to the isolation mass before onset of migration. 
Hence, the maximum mass $M_{\rm max}$ is roughly equal to $M_{\rm iso}$,
which is consistent with the value of $M_{\rm max}$ obtained in our results.

The orbital separations are $\sim 5-6 r_{\rm H}$, which means
that the planets are more packed than in setA.
The larger number and smaller separations of the final planets
than those in setA suggest that the final configuration could be unstable.
As will be shown below, in setB, the final planets start orbit crossing 
immediately after the removal of disk gas.

\subsection{Stability after Disk Gas Depletion}
\label{subsec:stability}

All the runs in setA with type-I migration and
setB without it end with multiple planets. 
Because eccentricity damping due to gravitational drag is so strong that
the systems of these multiple planets are orbitally stable
\citep{iwasaki02,kominami02}, although the minimum orbital separations
in the systems are rather small ($\sim 5-7 r_{\rm H}$).
However, disk gas should dissipate on timescales less than $10^7$ years.

In the gas-free case, it is predicted that these planets may become 
unstable on timescales of 
$t_{\rm cross} \sim 10^5 T_{\rm K} \sim 10^4$ years \citep{chambers_etal96}.
Note, however, that these timescales are for non-resonant planets.
In our results, the final multiple planets are usually captured in 
mean-motion resonances that generally stabilize the systems. 
Actually, \citet{terquem07} found that the final multiple planets,
which correspond to our results in runA1 - runA4, are stable 
on timescales much longer than $t_{\rm cross}$ even after
removal of disk gas.
We show similar results for runA1 - runA4 below, but also show that
the removal of disk gas makes the systems unstable for runB1 - runB4. 

To examine long term stability, we take out the planets
with masses larger than $0.01 M_{\oplus}$ from the final state of
the runs, and integrate their orbits, neglecting many other 
smaller-mass planets for saving computational cost.
As expected, we found that the system of runA1 is stable until
$2 \times 10^7 T_{\rm K}$ under the disk gas damping.
To study the stability after the disk gas removal, 
we re-start the calculation from the final state of runA1 
at $5 \times 10^6 T_{\rm K}$ without the disk gas damping.
We adiabatically adjust the orbital configuration 
to the gas-free condition, by decreasing $f_g$ as
\begin{equation}
f_g = \exp\Bigl(-\frac{t- 5 \times 10^6 T_{\rm K}}{10^4{\rm years}}\Bigr).
\end{equation}
The decay timescale of $10^4 {\rm years} (\sim 1.4 \times 10^5 T_{\rm K})$ 
is long enough for the adjustment.

At the end of runA1 (Fig.~\ref{fig:t_a_run1}), 
six planets are remain with separations of $5 - 9 r_{\rm H}$. 
The orbital evolution after the disk gas removal of runA1 
is shown in Fig.~\ref{fig:t_a_dis} (the left panel).
The figure shows that the eccentricities of the planets are 
kept less than 0.01 and the planets remain stable even after 
the gas removal.
Although the minimum orbital separation is about $5 r_{\rm H}$,
the resonant configuration stabilizes the system.

The other runs (runA2 - runA4) in this set with type-I migration
also show similar results: 
Even after the disk gas removal, 
orbital separations hardly change (the average orbital separation is 
$\overline{\Delta a} = 9.5 \pm 0.97 r_{\rm H}$) and 
most of the planets keep 
their commensurate relationships until the end of simulations. 
The average orbital eccentricity is $\overline e = 0.0086 \pm 0.0061$.
This is consistent with the result by \citet{terquem07}.

The stability for setB without type-I migration
is completely different.
At the end of runB1 (Fig.~\ref{fig:t_a_model59}), 
45 planets with masses larger than $0.01 M_{\oplus}$ remain with
the orbital separations of the planets are $5-6 r_{\rm H}$. 
Figure~\ref{fig:t_a_dis} (the right panel) shows the semimajor axis evolution 
of runB1 after the disk gas removal. 
Soon after the gas removal,
the eccentricities are pumped up and the planets start orbit crossing.
Note, however, that the planets do not exhibit global orbital crossing 
that the planets at $\sim 1$AU around solar-type stars 
exhibit after disk gas removal (e.g., \citealt{kominami02}). 
In terrestrial planet regions around M dwarfs ($\sim 0.1$AU),
physical radii of the planets $r_{\rm p}$ 
relative to their Hill radii $r_{\rm H}$ are larger than that in $\sim 1$AU
($r_{\rm p}/r_{\rm H} \propto M_*^{1/3} a^{-1}$).
Then, the eccentricities are pumped less highly and furthermore 
merging proceeds before the eccentricities are fully excited.
Thus, the planets collide with only neighboring planets. 
Finally, nine planets with moderate eccentricities ($\sim 0.08$) 
are formed. 
The mass of the largest planet is $0.65 M_{\oplus}$. 
All the commensurabilities are lost 
in the course of close encounters and
the final planets are not trapped in mean motion resonances at all.
The orbital separations are $\sim 20 r_{\rm H}$,
which are large enough to be dynamically isolated from each other
in the non-resonant configurations. 
Since relatively many planets ($\sim 40$) are formed 
with small orbital separations, 
all the runs in setB without type-I migration
exhibit orbital crossing and merging of the planets after
the disk gas removal, resulting in
$\sim 10-20$ planets with the average eccentricity of $0.055 \pm 0.020$
and the average orbital separation of $19 \pm 2.2 r_{\rm H}$ that lost 
commensurate relationships.
The average mass of the largest planet is $\overline M_{\rm max} = 0.50 \pm 0.097 M_{\oplus}$.
The non-resonant orbital configurations with wider separations
and larger eccentricities are the characteristic to  setB
without type-I migration (with 100 times reduced migration efficiency).
The semimajor axes of the planets are not concentrated 
to the regions near the disk inner edge, in contrast to
the results with type-I migration.
Thus, type-I migration efficiency in inner disk regions regulates
orbital configurations of close-in terrestrial planets.

\subsection{Dependence on Boundary and Initial Conditions}

Since we are concerned with planetesimal accretion near
the disk inner edge, we study the effects of general relativity
that is effective in the proximity of the host star
and reserved type-I migration torque that occurs near the edge.
We re-started the stability calculation in gas-free condition
for the runA1 - runA4, incorporating the relativistic effect 
directly into orbital integration. 
The detailed expression of post-Newtonian gravitational force 
from the host star is given in Appendix~\ref{ap:rel}.
Although the relativistic effect causes the precession of 
the perihelion of short-period planets, 
we found that the resonant relationships are not changed by 
the relativistic effect and 
the systems stayed in a stable state. 
\citet{terquem07} found that the tidal dissipation 
does not affect the stability as well as the relativity.

It is argued that inward protoplanet migration can be 
halted before reaching the inner cavity, because the tidal torque 
from the disk is reversed due to inverse pressure gradient
near the disk edge \citep{masset_etal06}.
Because the planets in the inverse torque regions gain
angular momentum from the disk gas, the inwardly migrating
planets in outer regions cannot push the inner planets into
the cavity, before the depletion of disk gas.
Substituting the component of the gas surface density gradient
at the inner edge in our model into $-q$ in Eq.~(\ref{eq:t_mig_2}),
we performed simulations in runC1 and runC2 (This effect was also investigated 
in \citet{terquem07}).
The orbital evolution for runC1 is shown in Fig.~\ref{fig:t_a_model53}.
Although qualitative evolution is almost the same as 
the result without the reversed torque, the number of final planets 
is fewer than that obtained in runA1 - runA4. 
Because the innermost planet cannot penetrate into the cavity,
the orbital separation between the innermost planet and 
the second innermost one is small ($\sim 3.8 r_{\rm H}$).
However, the right panel of Fig.~\ref{fig:t_a_model53} shows that
the planets remain orbitally stable after the disk gas removal,
because they keep being trapped in 5:4 and 9:8 resonances.

Around M dwarfs, the ice line is so close that significant amount of
icy protoplanets are quickly formed and migrate to inner regions.
The migrated icy protoplanets affect accretion of planets 
in terrestrial planet regions, in particular,
they regulate the mass of the largest planet in final state. 
So far, we have adopted the ice condensation factor as $\eta_{\rm ice} = 3$
in Eq.~(\ref{eq:solid}).
However, more enhanced $\eta_{\rm ice}$ was proposed by several authors.
\citet{stevenson88} proposed that sublimation of icy grains
that have migrated to inside the ice line
and the diffusion of the water vapor 
enhances the surface density of icy materials near the ice line
up to $\eta_{\rm ice} \sim 75$. 
Recent more detailed study \citep{ciesla06} showed $\eta_{\rm ice} \sim 10$.
To highlight the effect of migrated icy protoplanets, 
we performed calculations with 
\begin{eqnarray}
\eta_{\rm ice} = \Bigl \{
\begin{array}{ll}
1 & [r < 0.3 {\rm AU}]\\
14 & [0.3 {\rm AU} < r].
\end{array}
\Bigr. 
\end{eqnarray}

The result including the effect of type-I migration (runD1)
is shown in Figure~\ref{fig:t_a_model37}.
It is qualitatively similar to runA1 with $\eta_{\rm ice}=3$ at $r > 0.3$AU:
several planets are formed near the disk inner edge 
trapped in mutual mean motion resonances before the disk removal. 
Even if disk gas is removed, the final planets are stable.
We also did three additional runs with the same setting
(runD2 - runD4) and they show similar results.
However, the average mass of the largest planet is 
$\overline M_{\rm max}=3.6 \pm 0.29 M_{\oplus}$, which is
significantly larger than that with $\eta_{\rm ice}=3$ (setA).
This difference means that the planets mostly consist of
migrated icy protoplanets.
Because orbital separations must be greater than $\sim 5 r_{\rm H}$
for the planets to be stable and
$r_{\rm H}$ is proportional to $M^{1/3}$,
the average number of final planets ($\overline N=4.0 \pm0.71$) 
is smaller than that in setA.

We also performed runs with $\eta_{\rm ice}=14$ that did not include 
type-I migration (setE; runE1 - runE4).	
The result for runE1 is shown in Fig.~\ref{fig:t_a_model38}.
Before the disk gas removal, 
the results in setE are similar to those in setB with $\eta_{\rm ice}=3$, 
except for the final maximum mass 
($\overline M_{\rm max}=0.91 \pm 0.064 M_{\oplus}$) 
and number ($\overline N=27\pm2.3$).
In setB, resonant trapping was so efficient that
$M_{\rm max}$ was no other than $M_{\rm crit,mig}$.
In setE, the inner protoplanets cannot halt the migrated protoplanets
from outer regions, total mass of which is one order larger than
the total mass of inner planets, and mergers and rearrangement 
occur in the inner regions. 
This results in the larger $M_{\rm max}$ and smaller $\overline N$
in setE than in setB.

So far, we have been using $\Sigma_g \propto r^{-1.5}$.
We also did several simulations with less steep radial gradient,
$\Sigma_g \propto r^{-0.5}$.
Because of weaker dependence of type-I migration timescale on $r$
corresponding to the weaker $r$-dependence of $\Sigma_g$,
we found that a few protoplanets that are trapped 
by each other migrate together, which \citet{McNeil_etal05} called a ``convoy.''
But, we also found that this feature does not affect
the final orbital configurations of close-in planets before disk gas
removal and their stability after the disk gas removal.
The final orbital configurations
depend on only type-I migration speed around 0.1AU, because it 
determines the efficiency of resonant trapping.

\subsection{Composition and Habitability}
\label{subsec:comp}

As stated in section 1, 
delivery of icy planetesimals from the regions beyond 
the ice line is one of likely sources for the $\rm{H_2O}$-water 
on planetary surface (\citealt{morbidelli_etal00}, \citealt{robert01}). 
Assuming this scenario, 
\citet{raymond_etal07} suggested through \textit{N}-body simulation 
neglecting type-I migration
that the planets in HZs around M dwarf stars are likely to be dry, 
since radial mixing is inefficient in the lower-mass disks.

Figure~\ref{fig:comp} shows $\rm{H_2O}$-water mass fraction of the final planets
in runA1 (the left panel) and runB1 (the right panel), 
using the following simple prescription
for components of planetesimals that originated at $r$:
\begin{eqnarray}
\frac{M_{{\rm water}}}{M}=\frac{\eta_{\rm ice} -1}{\eta_{\rm ice}} = \Bigl \{
\begin{array}{ll}
0 & [r < r_{\rm ice}]\\
0.67 & [r > r_{\rm ice}].
\end{array}
\Bigr. 
\end{eqnarray}
Note that the initial mass beyond the ice line make up 41\% 
of total mass in our calculation range, following the MMSN model (Eq.~[\ref{eq:solid}]).
The shaded regions in Fig.~\ref{fig:comp} represent 
analytically estimated HZ (\citealt{kasting_etal93}, 
\citealt{selsis_etal07}). 
Note that the positional relationship between the disk inner edge and 
the HZ is not exact because the locations of disk inner edge are uncertain.
In both runA1 and runB1, significant amount of water was
delivered by planetary migration.
As a result, the final planets are considerably ``wet''
except for the innermost planets that are
shielded by outer planets.
Thus, water delivery to the HZ is rather efficient around M dwarfs 
and the terrestrial planets would be rich in water. 
Even if the effect of type-I migration is fully retarded, 
water-rich protoplanets migrate inward by the eccentricity damping
due to gravitational drag. 
The right panel of Fig.~\ref{fig:comp} shows that
water-rich planets in HZs may be usually formed for relatively
slow migration.

\subsection{Dependence on Disk Mass}

So far, we have adopted $f_g = 1$ for protoplanetary disk in Eq.~(\ref{eq:Sigma}) which is relatively large for the disks around 
$M_* = 0.2 M_{\odot}$ stars. 
The reduced computational cost due to the high $f_g$ allows us
to carry out large enough number of runs 
for the statistical arguments.  
Here we discuss how the results can be changed if
we consider less-massive disks for M dwarfs with 
$f_g \sim 0.2$, which may be averaged values
(\citet{raymond_etal07} adopted $f_g \simeq 0.15$).
As described above, the results of \textit{N}-body simulations
are explained well by using the timescales 
derived in section 2.3.
We discuss results of planetary formation in less massive disk
by applying the timescales for smaller values of $f_g$.

We found that the migration speed regulates 
final orbital configurations of the close-in terrestrial planets.
The migration is slower in less massive disks.
According to Eqs.~(\ref{eq:t_mig}) and (\ref{eq:damp_a_2}), 
both the timescales of type-I migration and the migration 
induced from eccentricity damping are 
inversely proportional to the disk gas scaling factor $f_g$.
Thus, in the less massive disks, the final planets 
tend to have relatively large separations 
in non-resonant orbits because of the slower migration.

We found that the terrestrial planets around M dwarfs 
are generally water-ice rich 
by the relatively fast migration of icy protoplanets
due to the relatively small radius ($\sim 0.3$AU) 
of the ice line.
From Eqs.~(\ref{eq:M_crit_mig}) and (\ref{eq:m_crit}) with
$f_g = f_d = 0.2$, the critical masses for retention
against type-I migration and the migration induced 
from eccentricity damping are given respectively by
\begin{eqnarray}
\label{eq:m_crit_mig_2}
&&M_{\rm crit, mig} \simeq 0.17 \Bigl(\frac{\eta_{\rm ice}}{3}\Bigr)^{3/4} \Bigl(\frac{f_d}{0.2}\Bigr)^{3/4} 
\Bigl(\frac{f_g}{0.2}\Bigr)^{-9/20} 
\Bigl(\frac{a}{0.3 \mathrm{AU}}\Bigr)^{-9/10} 
M_{\oplus},\\
\label{eq:m_crit_2}
&&M_{\rm crit,damp} \simeq 3 \Bigl(\frac{\eta_{\rm ice}}{3}\Bigr)^{3/4} \Bigl(\frac{f_d}{0.2}\Bigr)^{3/4} 
\Bigl(\frac{f_g}{0.2}\Bigr)^{-9/20} 
\Bigl(\frac{e}{0.01} \Bigr)^{-3/2} \Bigl(\frac{a}{0.3 \mathrm{AU}}\Bigr)^{-21/40} M_{\oplus}.
\end{eqnarray}
The isolation mass is (Eq.~[\ref{eq:m_iso}]) 
\begin{eqnarray}
\label{eq:m_iso_2}
M_{\rm iso} \simeq 0.24 \Bigl(\frac{\eta_{\rm ice}}{3}\Bigr)^{3/2} \Bigl(\frac{f_d}{0.2}\Bigr)^{3/2} 
\Bigl(\frac{a}{0.3 \mathrm{AU}}\Bigr)^{3/4} 
M_{\oplus}.
\end{eqnarray}
Comparison of these masses shows that protoplanets just outside the ice line 
almost grow to the isolation mass before starting migration. 
Substituting the isolation mass into Eq.~(\ref{eq:t_acc}), 
we obtain the accretion time as
\begin{eqnarray}
t_{\rm acc} \simeq 8.0 \times 10^4 \Bigl(\frac{\eta_{\rm ice}}{3}\Bigr)^{-1/2} 
\Bigl(\frac{f_d}{0.2}\Bigr)^{-1/2} \Bigl(\frac{f_g}{0.2}\Bigr)^{-2/5} 
\Bigl(\frac{a}{0.3 {\rm AU}}\Bigr)^{59/20}
\,{\rm years}.
\end{eqnarray}
Similarly, substituting Eq.~(\ref{eq:m_iso_2}) into Eqs.~(\ref{eq:t_mig}) and (\ref{eq:damp_a_2}), 
we obtain the migration timescales as
\begin{eqnarray}
&&t_{\rm mig} \simeq 2.4 \times 10^4 \Bigl(\frac{\eta_{\rm ice}}{3}\Bigr)^{-3/2} \Bigl(\frac{f_d}{0.2}\Bigr)^{-3/2} 
\Bigl(\frac{f_g}{0.2}\Bigr)^{-1}
\Bigl(\frac{a}{0.3 \, \mathrm{AU}}\Bigr)^{3/4} 
\, \mathrm{years}, \\
&&t_{{\rm damp},a} \simeq 1.3 \times 10^6 
\Bigl(\frac{\eta_{\rm ice}}{3}\Bigr)^{-3/2} \Bigl(\frac{f_d}{0.2}\Bigr)^{-3/2} 
\Bigl(\frac{f_g}{0.2}\Bigr)^{-1} \Bigl(\frac{e}{0.01}\Bigr)^{-2}
\Bigl(\frac{a}{0.3 \, \mathrm{AU}}\Bigr)^{5/4} 
\, \mathrm{years},
\end{eqnarray}
where $t_{\rm mig}$ is the type-I migration timescale and $t_{{\rm damp},a}$ is the the timescale of the $e$-damping induced migration.
Both $t_{\rm mig}$ and $t_{\rm acc}$ are significantly smaller
than disk lifetimes of $\sim 10^6-10^7$ years even for 
$f_g = f_d \sim 0.2$.
Therefore, our finding that the close-in terrestrial planets around M dwarfs are rather ``wet'' is still valid for the averaged-mass disks with $f_g = f_d \sim 0.2$ around
$M_* \sim 0.2 M_{\odot}$.

Note, however, that the upper limit of migration timescale, 
$t_{{\rm damp},a}$, is comparable to the disk lifetimes.
If the type-I migration timescale is elongated by a factor of
more than 100, $t_{\rm mig}$ is also 
shorter than or comparable to the disk lifetimes.
Then, the transfer of water/icy materials by migrations of
protoplanets are not efficient enough.
If the migration barrier 
near the ice line 
(\citealt{kretke07}, \citealt{ida08b}, \citealt{kretke_etal09})
is effective also for disks around M dwarfs, 
the transfer is inefficient irrespective of
the reduction factor of type-I migration.
Since \citet{raymond_etal07} showed that 
the transfer of water/icy materials by scattering is
also inefficient, the planets in HZs
are not always rich in water-ice in these cases.
Thus, whether the planets in HZs around M dwarfs are
habitable or not may be strongly regulated by
efficiency of type-I migration.  

\section{CONCLUSIONS AND DISCUSSION}
\label{sec:conc}

We have investigated accretion of terrestrial planets from planetesimals 
around M dwarf stars through a set of \textit{N}-body simulations, including
the effects of disk gas. 
In general, accretion of terrestrial planets have two stages:
runway/oligarchic growth and following long-term giant impacts.
Our simulations cover all the stages from initial 5,000 planetesimals
to the final planets that are stable for long time after possible giant impacts,
fully including gravitational interactions of all the bodies. 

Since M dwarfs are fainter than solar-type stars,
both the HZs and ice lines are located in the proximity of central stars.
Due to the proximity, accretion of terrestrial planets have
different features around M dwarf than around solar-type stars that are caused
by the three factors:
\begin{list}{}{}
\item (a) the effective damping by disk gas due to high gas density in inner regions,
\item (b) the influence from inner protoplanets that 
          have migrated toward the disk inner edge,
\item (c) the influence from outer icy protoplanets that 
          migrate into the terrestrial planet regions.
\end{list}

Regarding factor (a), it is noticed that
higher disk gas density due to the proximity overwhelms
expected smaller disk and planetary isolation masses around M dwarfs.
Around M dwarfs, the disk mass and consequently, isolation mass of protoplanets may be 
generally smaller than those around solar-type stars.
However, due to the higher disk gas density,
planet-disk tidal interactions, that is, eccentricity (and inclination) 
damping and type-I migration are more efficient than those at $\sim 1$AU
around solar type stars.  
Furthermore, accretion timescale in such regions is much shorter than disk lifetime
($\sim 10^6-10^7 {\rm yrs}$).
As a result, the eccentricity damping and type-I migration play important
roles in architecture of terrestrial planets around M dwarfs.
To highlight this effect, for \textit{N}-body simulations,
we adopted disks comparable to MMSN that
may be relatively massive among disks around M dwarfs.
Even in the case without type-I migration, the migration induced
from eccentricity damping, 
which is 100 times slower than type-I migration, 
is still fast enough to bring protoplanets
into the terrestrial planet regions for 
such disks. 
The dominance of the effects of disk gas was discussed with
analytical arguments (section 2.3 and 3.5).

Regarding factor (b), 
resonant trapping plays an important role.
Because of the efficient type-I migration, 
many protoplanets migrate 
toward the disk inner edge and accumulate there, usually trapped in 
mutual mean-motion resonances.
We set the disk inner edge to be 0.05AU, while 
the HZ is around $\sim 0.1$AU, which is only a factor 2 difference.
In the slow migration case, in which resonant trapping
is so efficient, trapped protoplanets line up through 
almost all the simulation 
regions (0.05-0.4AU), while with full migration speed
predicted by the linear theory,
trapping is not so efficient that coagulation often occurs and
the final planets tend to be concentrated in inner regions.
In the former case, about 40 planets remain in the resonances
before the disk gas removal.
After the disk gas is removed, the orbits of the planets
become unstable and giant impacts occur.
As a result, widely-spaced ($\sim 20 r_{\rm H}$), non-resonant, 
multiple planets are formed with relatively high eccentricities ($\sim 0.05$) 
between disk inner edge and outer region.
On the other hand, in the full migration case,
the trapped planets, the number of which is about 5, are stable even after the disk gas removal
and closely-packed ($\sim 5-10r_{\rm H}$), resonant planets
remain in the proximity of the disk inner edge with 
low eccentricities ($\la 0.01$).

Therefore, we conclude that the migration speed is a key factor for
final orbital configuration of close-in terrestrial planets
around M dwarfs.
The close-in planets that on-going radial velocity surveys have
discovered may support the slow migration.
The three-planet system around Gl~581 with $M_* \simeq 0.3M_{\odot}$
is composed of planet b ($M_{\rm p} \sin i =16M_{\oplus}, a=0.041$AU), 
c ($5 M_{\oplus}, 0.073$AU), 
and d ($7.5 M_{\oplus}, 0.25$AU).
They are widely-spaced ($\Delta a \sim 21 - 47 r_{\rm H}$) non-resonant planets that are consistent with
our slow migration model, although this system
was probably formed from heavier disk than the MMSN disk and may 
need the enhancement of surface density of icy materials near the ice line.
Around solar-type stars, the radius of the ice line is larger, so the factor (c) may not be applied. 
However, the dependence of final configuration of close-in terrestrial planets on the migration speed 
can be applied to solar-type stars. 
The three-planet system around a K dwarf HD~40307 with $M_* \simeq 0.77 M_{\odot}$ consists of planet 
b ($M_{\rm p} \sin i =4.2M_{\oplus}, a=0.047$AU), c ($6.9 M_{\oplus}, 0.081$AU), 
and d ($9.2 M_{\oplus}, 0.13$AU) \citep{mayor_etal08}. 
They are also widely-spaced ($\Delta a \sim 17 - 20 r_{\rm H}$), non-resonant planets.
These configuration is explained by setB in our calculation.

Note that the final state could be altered by other effects which 
we did not address.
We will discuss the effects of random torques exerted 
by strong disk turbulence due to Magneto-Rotational Instability in a
separate paper.
Two mechanisms have been proposed 
which lead to the excitation of eccentricities of the bodies 
and additional collisions between them: 
(i) during type-I migration of a gas giant planet, 
its mean motion resonances (mainly 2:1 resonance) sweep 
the terrestrial planet region (e.g., \citealt{zhou05}),
or (b) during the dispersal of the gas disk, secular 
resonances caused by the gas giant planet and the disk 
sweep through the inner orbits
(e.g., \citealt{nagasawa_etal05}). 
However, since gas giants are generally rare around M dwarfs, 
these mechanisms do not play an important role for terrestrial
planets around M dwarfs.

Factor (c) is caused by the smaller radius of the ice line
around M dwarfs.
Less efficient planetesimal accretion due to
the smaller disk surface density is overwhelmed by the faster accretion
due to smaller radii of icy regions ($\ga 0.3$AU) than those around
solar-type stars.
Combined with factor (a), migration of the icy protoplanets
into the terrestrial planet regions is so efficient that
the largest final planets have significant mass of icy components
except for the innermost one that is shielded from impacts
of icy protoplanets.
In the case of enhanced surface density of ice near the ice line,
the largest final planets are mostly composed of ice but not rocks.
It is interesting that the estimated bulk density 
for a $20M_{\oplus}$ transiting planet, 
GJ~436b, around a M dwarf
is consistent with water ice (\citealt{gillon_etal07}, \citealt{deming_etal07}).
Although the mass and number of final planets are affected by
this factor as well as the inner boundary conditions for type-I migration
and radial gradient of $\Sigma_g$ (see section 3.3),
the stability of the resonantly trapped planets after the disk gas removal, 
that is, orbital separation, resonant or non-resonant configuration, 
eccentricities of final planets are determined only by migration speed.

For the disks with $f_g$ that is several times smaller than
that of our fiducial model ($f_g=1$), which may be typical disks 
around M dwarfs, 
accretion and migration timescales are still much shorter than
disk lifetime, so the formed close-in planets are abundant in water-ice.
However, if type-I migration speed is more than 100 times slower than that
predicted by the linear theory or
the migration is trapped near the ice line
(\citealt{kretke07}, \citealt{ida08b}, \citealt{kretke_etal09}), 
the final close-in terrestrial planets would be rocky
due to the inefficient water delivery by the migration,
because radial mixing of planetesimals is also inefficient
around M dwarfs \citep{raymond_etal07}.
As \citet{lissauer07} pointed out, the ice line evolves
during relatively long pre-main sequence phase of M dwarfs.
We need to carry out detailed \textit{N}-body simulations 
in low-mass disks, including the evolution of the ice line
in a future work, 
to clarify the details on how wet are the planets in HZs around M dwarfs.

Our results around M dwarfs and calculations by \citet{terquem07}
around G dwarfs suggest that existence of close-in relatively-large terrestrial
planets are robust. 
However, our solar system does not have any close-in planet
inside 0.4 AU and large fraction of extrasolar planetary systems may not have
the close-in super-Earths either. 
We will address this issue elsewhere.

Our conclusion is that the migration speed determines
diversity of final orbital configuration of close-in terrestrial planets
around M dwarfs through the stability of the planets trapped in
mutual mean-motion resonances, so we will study more detailed
dependence on the migration speed as well as the dependence
on conditions of inner disk regions in a separate paper.
Around M dwarfs, these planets could be in HZs. Characteristics of 
habitable planets around M dwarfs are significantly affected by the details of these 
formation mechanisms. Although these close-in planets are well inside HZs for 
solar-type stars, their formation mechanisms may be similar.
Future observations of close-in terrestrial planets around M dwarfs as well as solar-type stars
by radial velocity surveys from ground
and transit surveys from space will constrain the migration efficiency
and the quantitative features of inner disk regions.

\vspace{1em} 
\noindent ACKNOWLEDGMENTS. 

We thank the anonymous referees for useful comments.
Numerical computations were in part carried out on GRAPE 
system at Center for Computational Astrophysics, CfCA, of National Astronomical Observatory of Japan.
This work was supported by Grant-in-Aid for JSPS Fellows 20008528.

\appendix

\section{EXPRESSION FOR AERODYNAMICAL GAS DRAG FORCE}
\label{ap:aero}
The aerodynamic gas drag force per unit mass is \citep{adachi_etal76}
\begin{eqnarray}
\textbf{\textit{F}}_{\rm aero} = - \frac{1}{2M} C_{\rm D} \pi r_{\rm p}^2 \rho_g \Delta u \Delta \textbf{\textit{u}},
\label{eq:f_aero}
\label{eq:aero}
\end{eqnarray}
where $C_{\rm D}=0.5$ is the gas drag coefficient, $r_{\rm p}$ is the physical radius of the body and $M$ is its mass. 
The density of disk gas $\rho_g$ is \citep{hayashi81}
\begin{equation}
\rho_g = 2.0 \times 10^{-9} f_g \Bigl(\frac{r}{1\mathrm{AU}}\Bigr)^{-11/4} \,\mathrm{g\, cm}^{-3},
\label{eq:eta}
\end{equation}
where $\Delta \textbf{\textit{u}}$ is the relative velocity of the body to the disk gas. 
Due to pressure gradient the velocity of disk gas 
$v_{\rm gas}$ 
is smaller than Kepler velocity $v_{\rm K}$ by a fraction \citep{adachi_etal76}
\begin{eqnarray}
\eta \simeq \frac{v_{\rm K}-v_{\rm gas}}{v_{\rm K}} = 1.8 \times 10^{-3} \Bigl(\frac{r}{1\mathrm{AU}}\Bigr)^{1/2},
\end{eqnarray}
where the temperature distribution of an optically thin disk given by Eq.~(\ref{eq:temp}) is used.

\section{EXPRESSION FOR GRAVITATIONAL GAS DRAG FORCE}
\label{ap:grav}
\citet{tanaka_etal02} and \citet{tanaka04} derived the damping forces exerted on the planet, through three-dimensional linear 
calculation,
\begin{eqnarray}
\textit{F}_{{\rm damp},r} 
   & = & \Bigl(\frac{M}{M_*}\Bigr) \Bigl(\frac{v_{\rm K}}{c_s}\Bigr)^4 
    \Bigl(\frac{\Sigma_g r^2}{M_*}\Bigr) 
    \Omega (2A ^{c}_{r}[v_{\theta} - r \Omega] + A ^{s}_{r} v_r) \label{eq:3-1} \\
\textit{F}_{{\rm damp}, \theta} 
   & = & \Bigl(\frac{M}{M_*}\Bigr) \Bigl(\frac{v_{\rm K}}{c_s}\Bigr)^4 
   \Bigl(\frac{\Sigma_g r^2}{M_*}\Bigr)
   \Omega (2A ^{c}_{\theta}[v_{\theta} - r \Omega] + A ^{s}_{\theta} v_r) \label{eq:3-2}  \\
\textit{F}_{{\rm damp},z} 
   & = & \Bigl(\frac{M}{M_*}\Bigr) \Bigl(\frac{v_{\rm K}}{c_s}\Bigr)^4 
   \Bigl(\frac{\Sigma_g r^2}{M_*}\Bigr) \Omega (A ^{c}_{z} v_z + A ^{s}_{z} z \Omega ) \label{eq:3-3} \\
\textit{F}_{{\rm mig},r} & = & 0 \label{eq:3-1-2} \\
\textit{F}_{{\rm mig}, \theta} 
   & = &  - 2.17\Bigl(\frac{M}{M_*}\Bigr) \Bigl(\frac{v_{\rm K}}{c_s}\Bigr)^2 
   \Bigl(\frac{\Sigma_g r^2}{M_*}\Bigr) \Omega v_{\rm K} 
   \label{eq:3-2-2} \\
\textit{F}_{{\rm mig},z} & = & 0, \label{eq:3-3-2}
\end{eqnarray}
where $\textbf{\textit{F}}_{\rm damp}$ is the specific damping force for $e$ and $i$, $\textbf{\textit{F}}_{\rm mig}$ 
is the specific damping force for $a$ and $\Omega$ is the Keplerian angular velocity. The coefficients are given by 
\begin{eqnarray}
A^{c}_{r} = 0.057  & & A^{s}_{r} = 0.176 \nonumber \\
A^{c}_{\theta} = -0.868  & & A^{s}_{\theta} = 0.325 \nonumber \\
A^{c}_{z} = -1.088  & & A^{s}_{z} = -0.871 .\nonumber  
\end{eqnarray}

\section{EXPRESSION FOR POST-NEWTONIAN GRAVITY FROM THE HOST STAR}
\label{ap:rel}
The specific force induced by the post-Newtonian gravity from the host star, $\textbf{\textit{F}}_{\rm rel}$,
given in \citet{kidder95} is
\begin{eqnarray}
\textbf{\textit{F}}_{\rm rel} &=& - \frac{G(M_* + M)}{r^2 c^2} \nonumber \\
& & \times \{ [(1+3\mu)v^2 - 2(2+\mu) \frac{G(M_* + M)}{r} -\frac{3}{2} \mu \dot{r}^2] \frac{\textbf{\textit{r}}}{r}
-2(2 - \mu) \dot{r} \textbf{\textit{v}} \},
\end{eqnarray}
where $c$ is the light speed, $\textbf{\textit{v}}$ is the velocity vector, $v = |\textbf{\textit{v}}|$, 
$\dot{r} = dr/dt$, and $\mu \equiv M_* M/(M_* + M)^2$.

To the lowest order of the eccentricity, the precession rate of the periastron is expressed as \citep{mardling02}
\begin{eqnarray}
\frac{d \varpi}{dt} = \frac{3 n GM_*}{a c^2},
\end{eqnarray}
where $n$ is the mean motion. The post-Newtonian effect is important only to the short-period planets.

\clearpage

\begin{figure}
\begin{center}
\epsscale{1.0}
\plotone{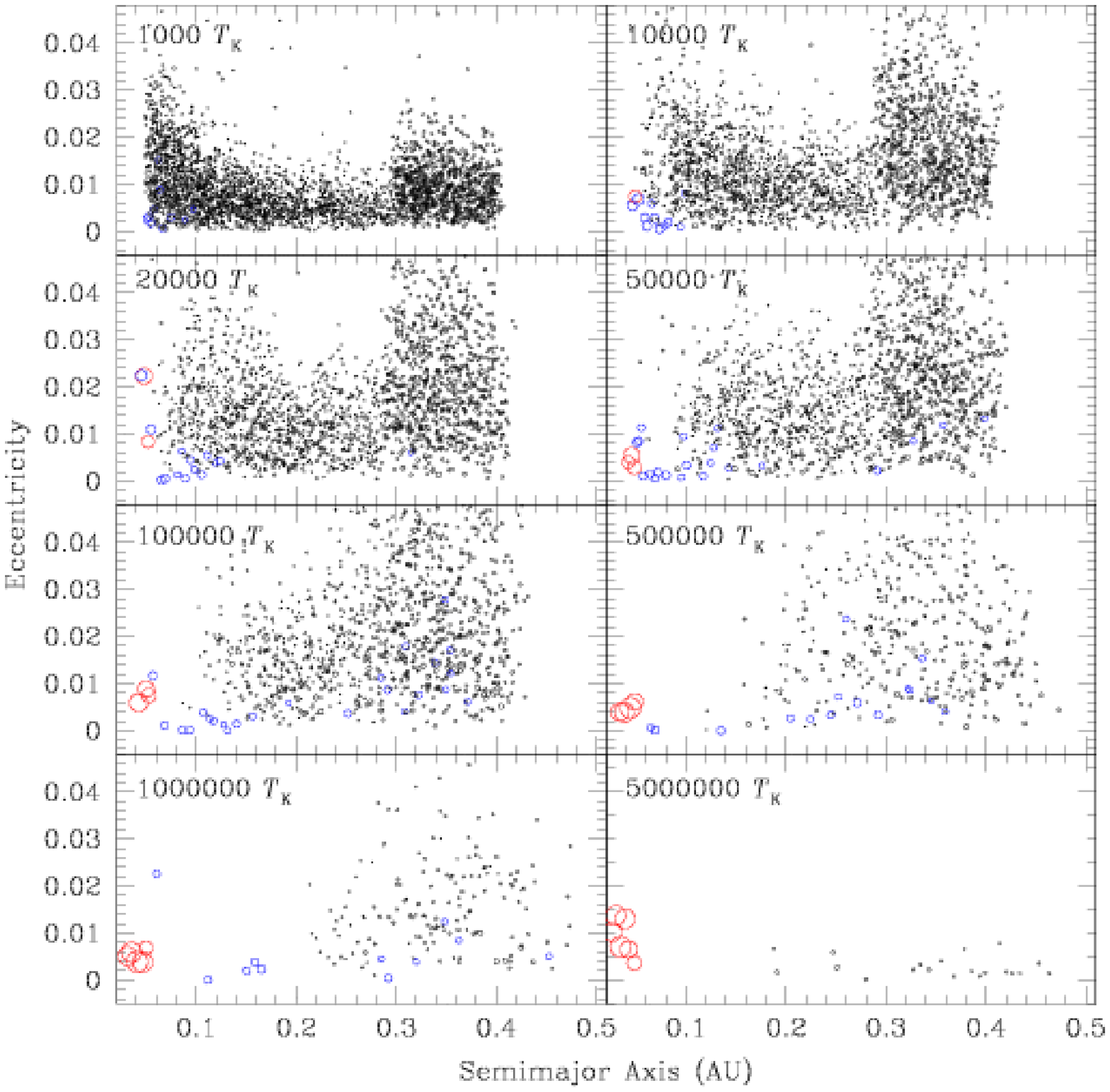}
\caption{Time evolution of a system on the $a-e$ plane. The circles represent bodies and 
the radii of the circles are proportional to the physical radii of the bodies. 
Note that overlapping circles does not mean the planets actually overlap each other 
because the size of circles are expanded so as to be easy to see.
The system initially consists of 5000 
planetesimals. The numbers of bodies are 2970 ($1000 \textit{T}_{\rm K}$), 2144 ($10000 \textit{T}_{\rm K}$), 
1862 ($20000 \textit{T}_{\rm K}$), 1448 ($50000 \textit{T}_{\rm K}$), 1107 ($100000 \textit{T}_{\rm K}$), 
390 ($500000 \textit{T}_{\rm K}$), 170 ($1000000 \textit{T}_{\rm K}$), and 26 ($5000000 \textit{T}_{\rm K}$).
$\textit{T}_{\rm K}$ is Keplerian time at 0.1~AU around a 
$0.2 M_{\odot}$ star, which is $\sim 0.071 yr$.
In the electronic version, bodies with $M > 0.01 M_{\oplus}$, $M > 0.1 M_{\oplus}$, and $M > M_{\oplus}$ are 
evpressed with blue, red, and green circles respectively. }
\label{fig:snap_run1}
\end{center}
\end{figure}

\begin{figure}
\begin{center}
\epsscale{1.0}
\plotone{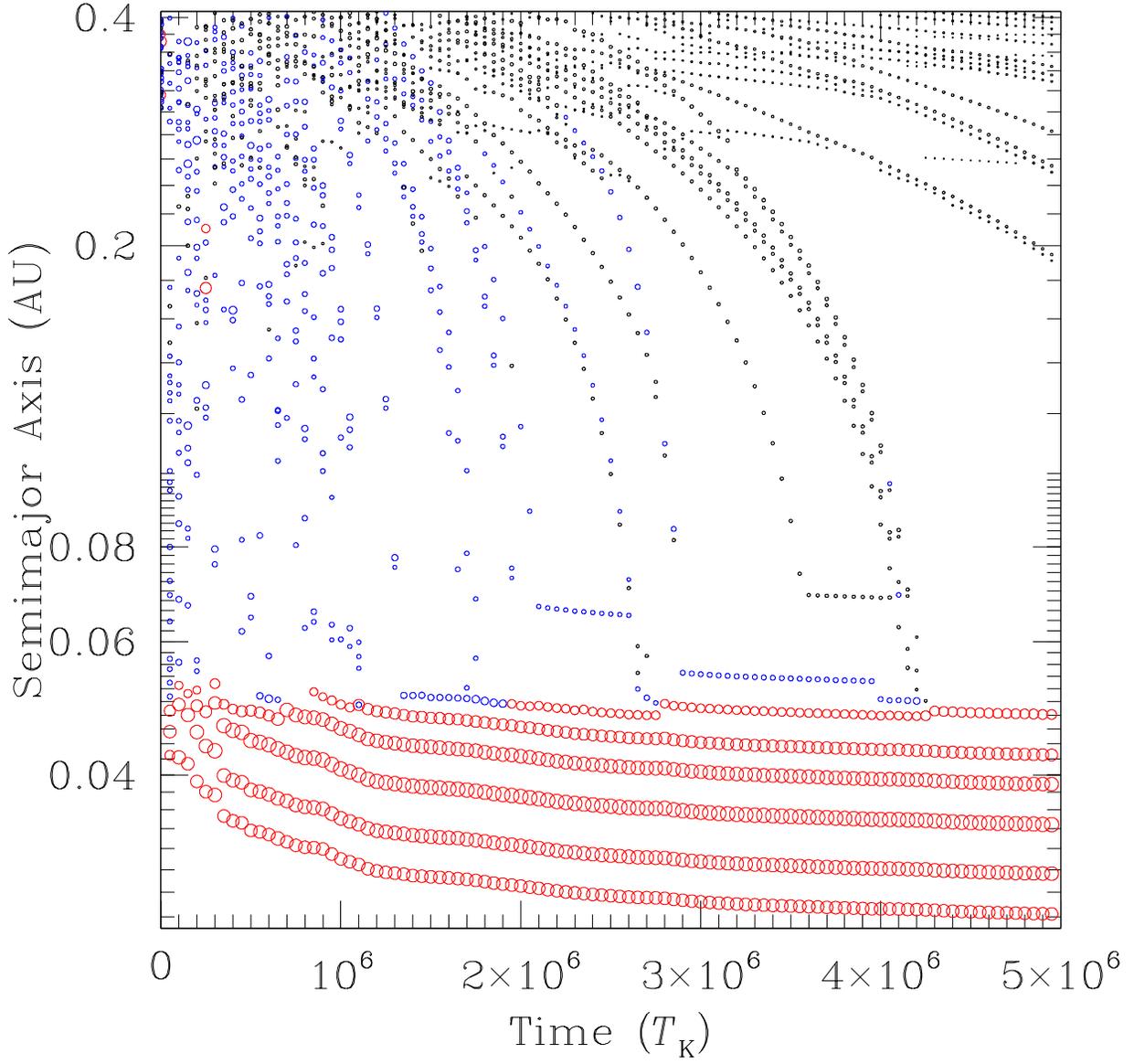}
\caption{Evolution of semimajor axes of massive planets.
At each time for runA1, 30 most massive planets are plotted. 
$\textit{T}_{\rm K}$ is Keplerian time at 0.1~AU around a 
$0.2 M_{\odot}$ star. 
The circles represent bodies and their radii are proportional to the physical radii of the bodies. 
}
\label{fig:t_a_run1}
\end{center}
\end{figure}

\begin{figure}
\begin{center}
\epsscale{1.0}
\plotone{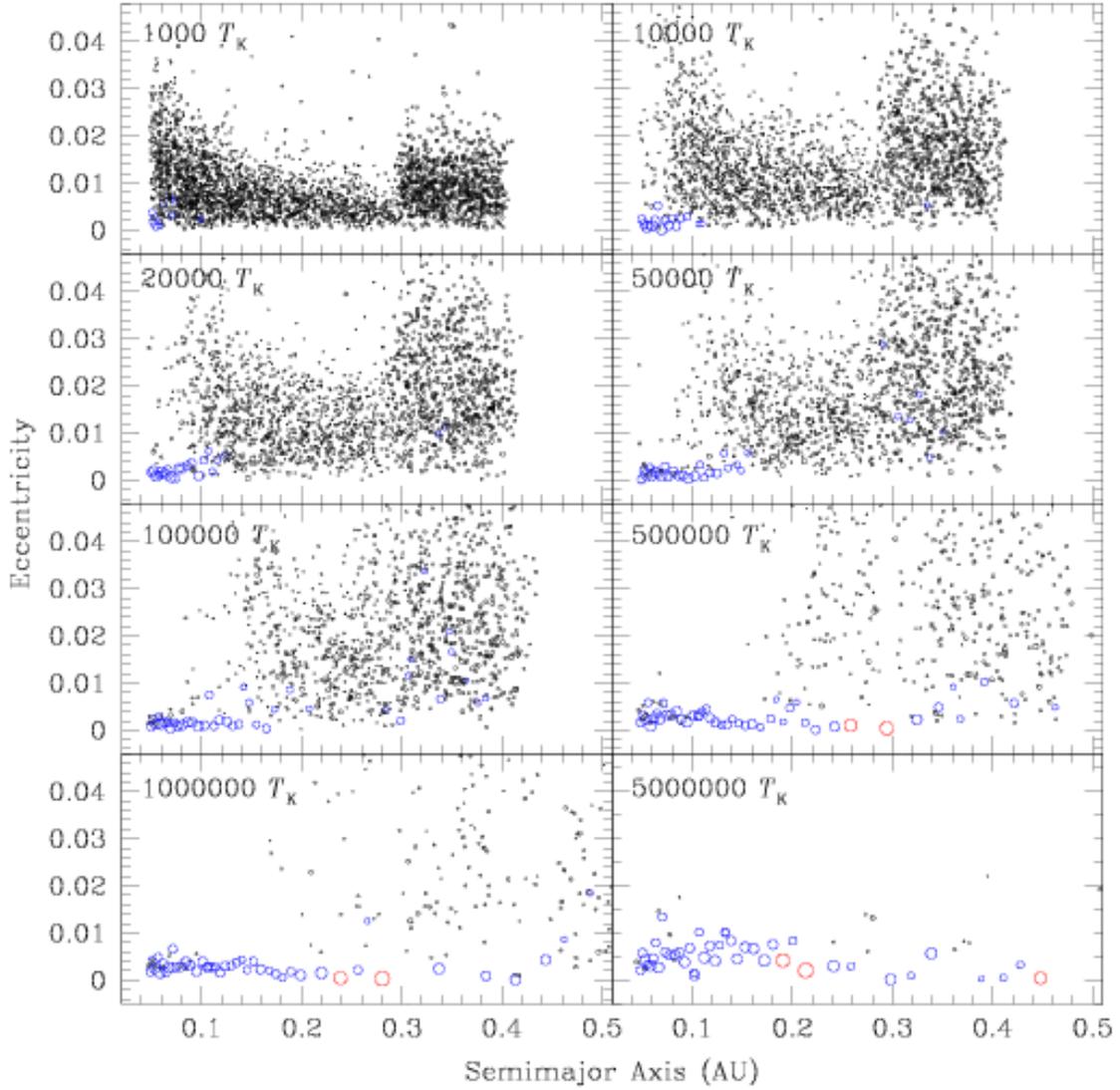}
\caption{The same as Fig.~\ref{fig:snap_run1} but the effect of type-I migration 
is not included. 
The system initially consists of 5000 
planetesimals. The numbers of bodies are 2961 ($1000 \textit{T}_{\rm K}$), 2158 ($10000 \textit{T}_{\rm K}$), 
1864 ($20000 \textit{T}_{\rm K}$), 1462 ($50000 \textit{T}_{\rm K}$), 1146 ($100000 \textit{T}_{\rm K}$), 
462 ($500000 \textit{T}_{\rm K}$), 231 ($1000000 \textit{T}_{\rm K}$), and 83 ($5000000 \textit{T}_{\rm K}$).}
\label{fig:snap_model59}
\end{center}
\end{figure}

\begin{figure}
\begin{center}
\epsscale{1.0}
\plotone{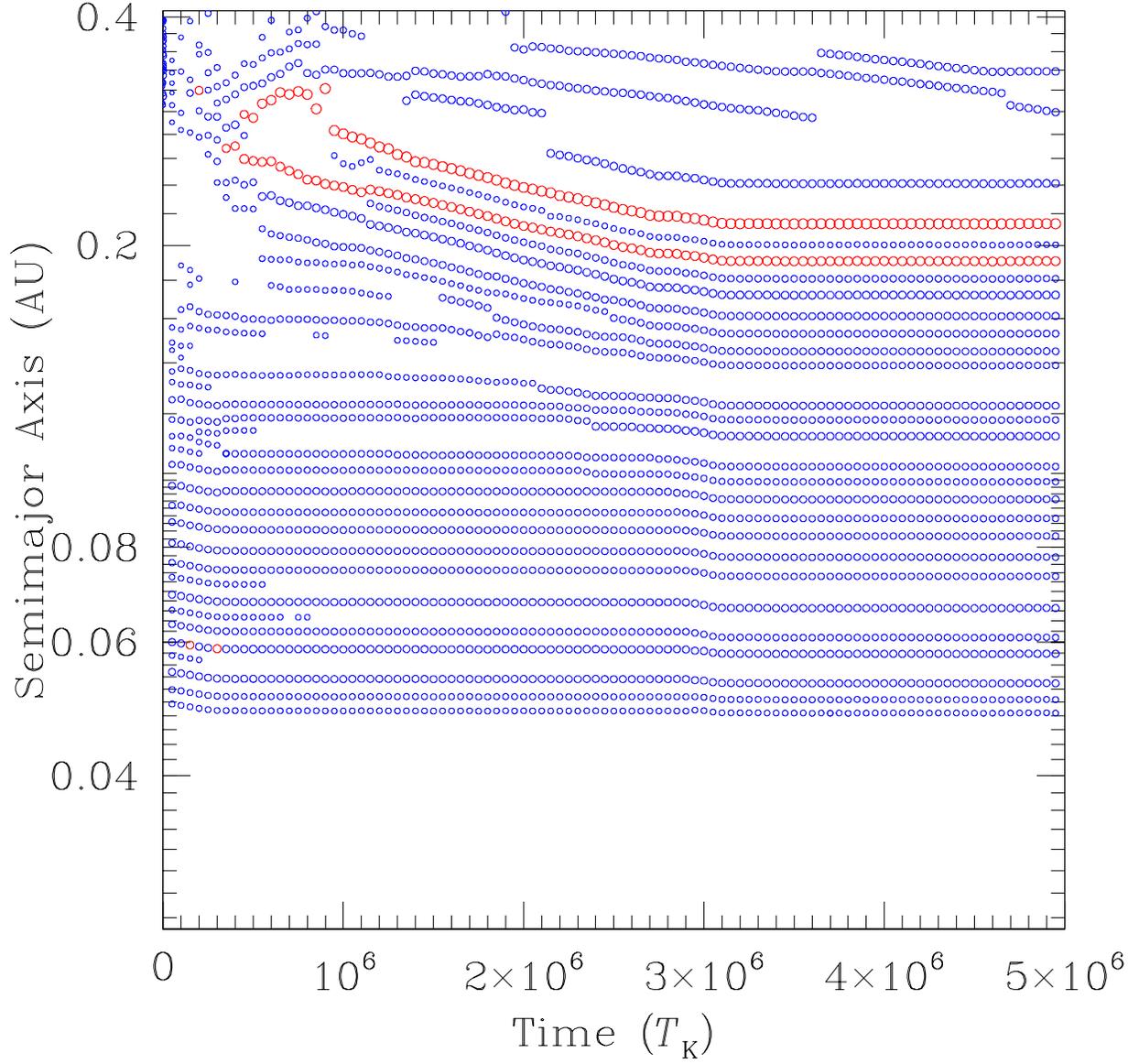}
\caption{The same plots as Fig.~\ref{fig:t_a_run1} for runB1. In this case, the effect of type-I migration in not 
included.}
\label{fig:t_a_model59}
\end{center}
\end{figure}

\begin{figure}
\begin{center}
\epsscale{1.0}
\plottwo{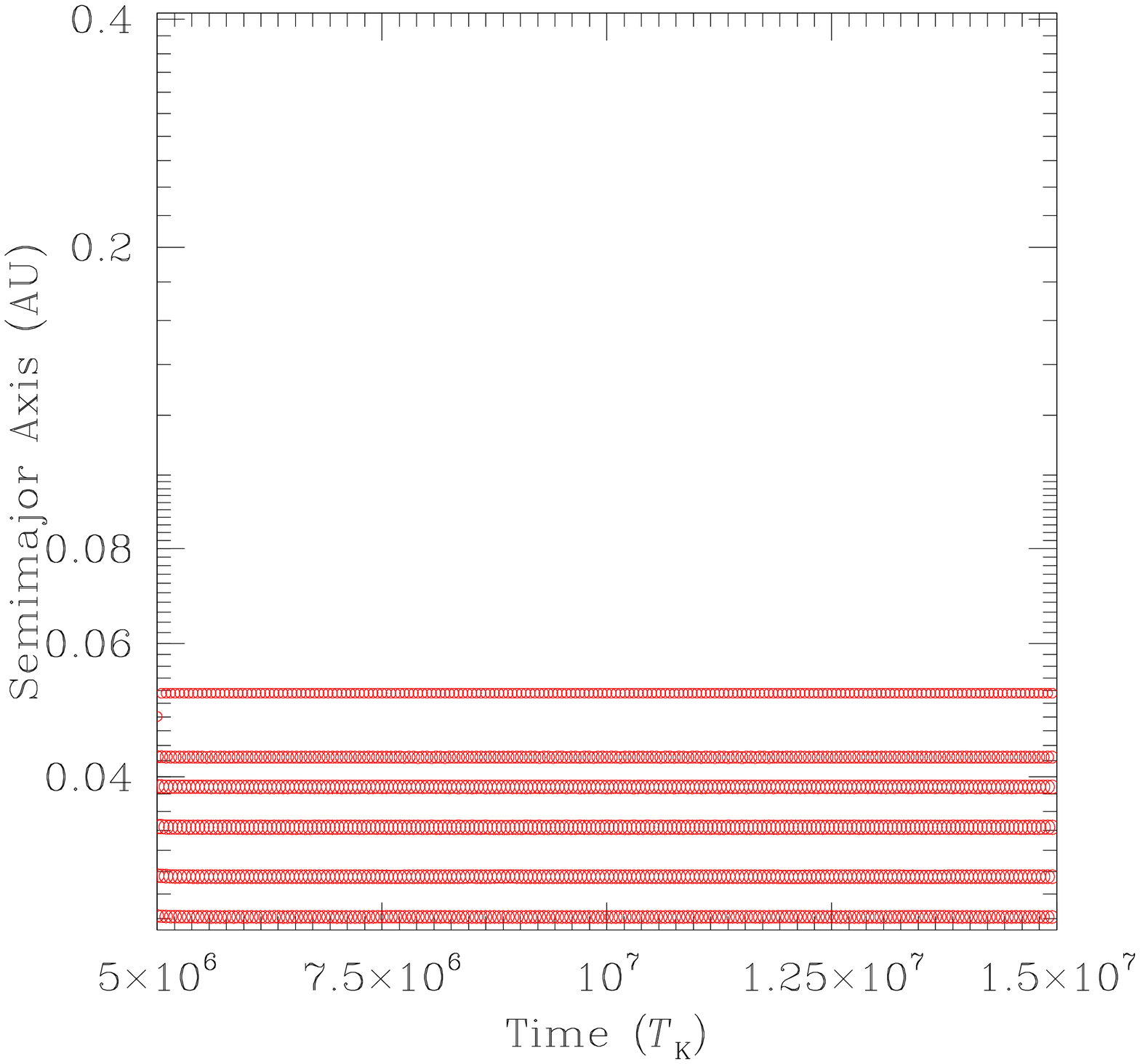}{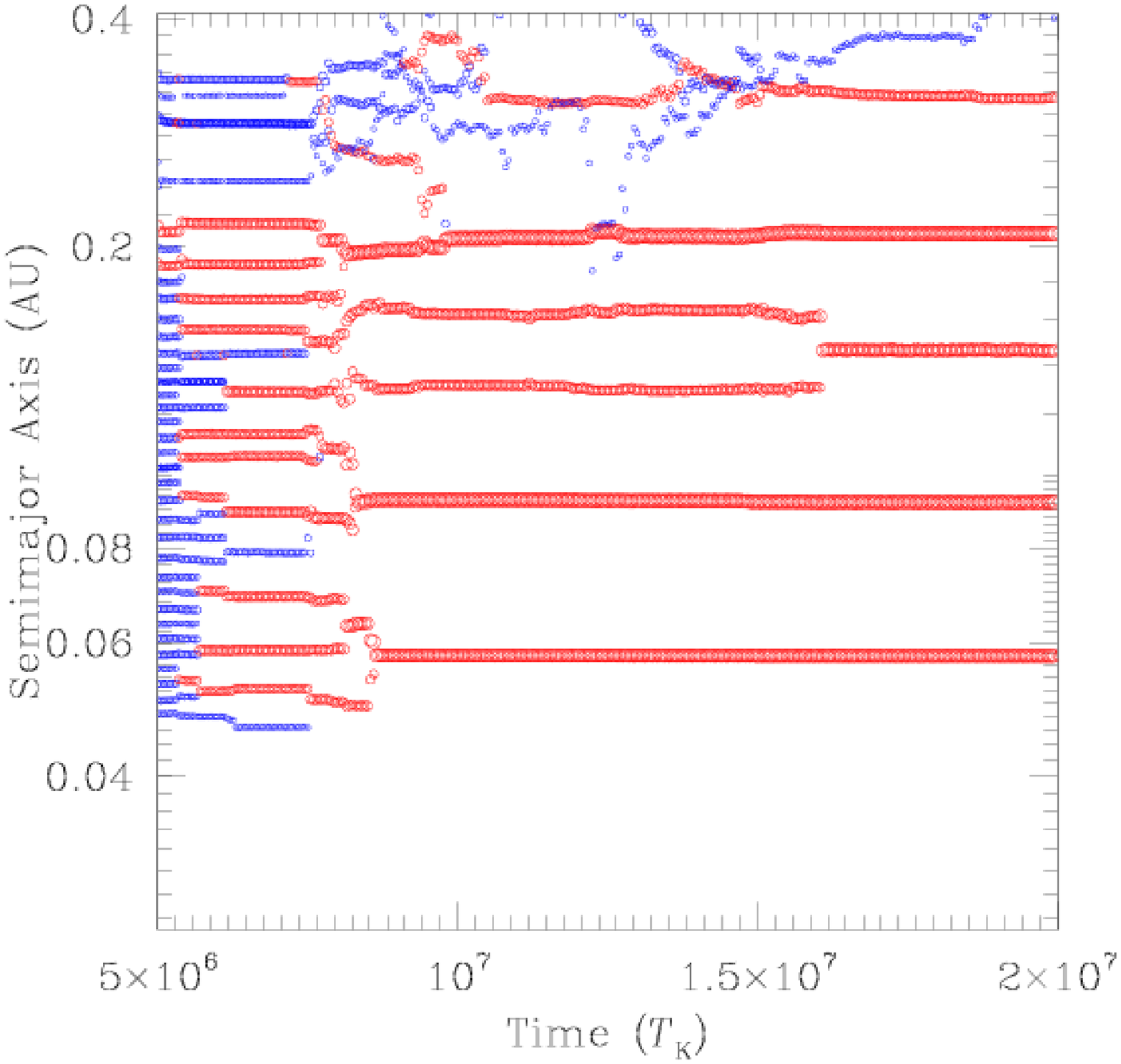}
\caption{Orbital evolution of successive simulation of runA1 (\textit{left}) and runB1 (\textit{right}) in a 
gas-free environment, neglecting small bodies.
In runB1, planets suffer giant impacts, which results in losing of commensurabilities.
}
\label{fig:t_a_dis}
\end{center}
\end{figure}

\begin{figure}
\begin{center}
\epsscale{1.0}
\plottwo{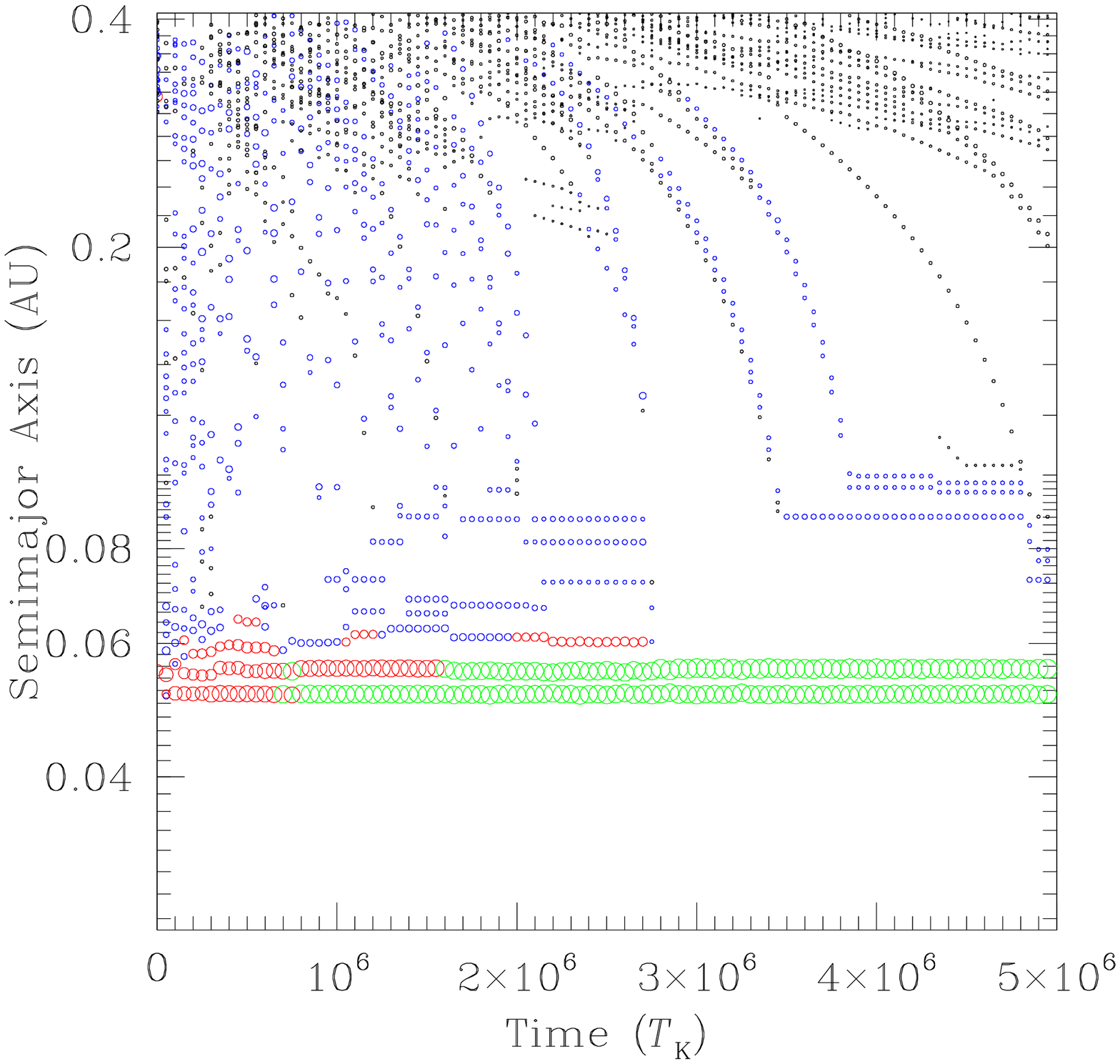}{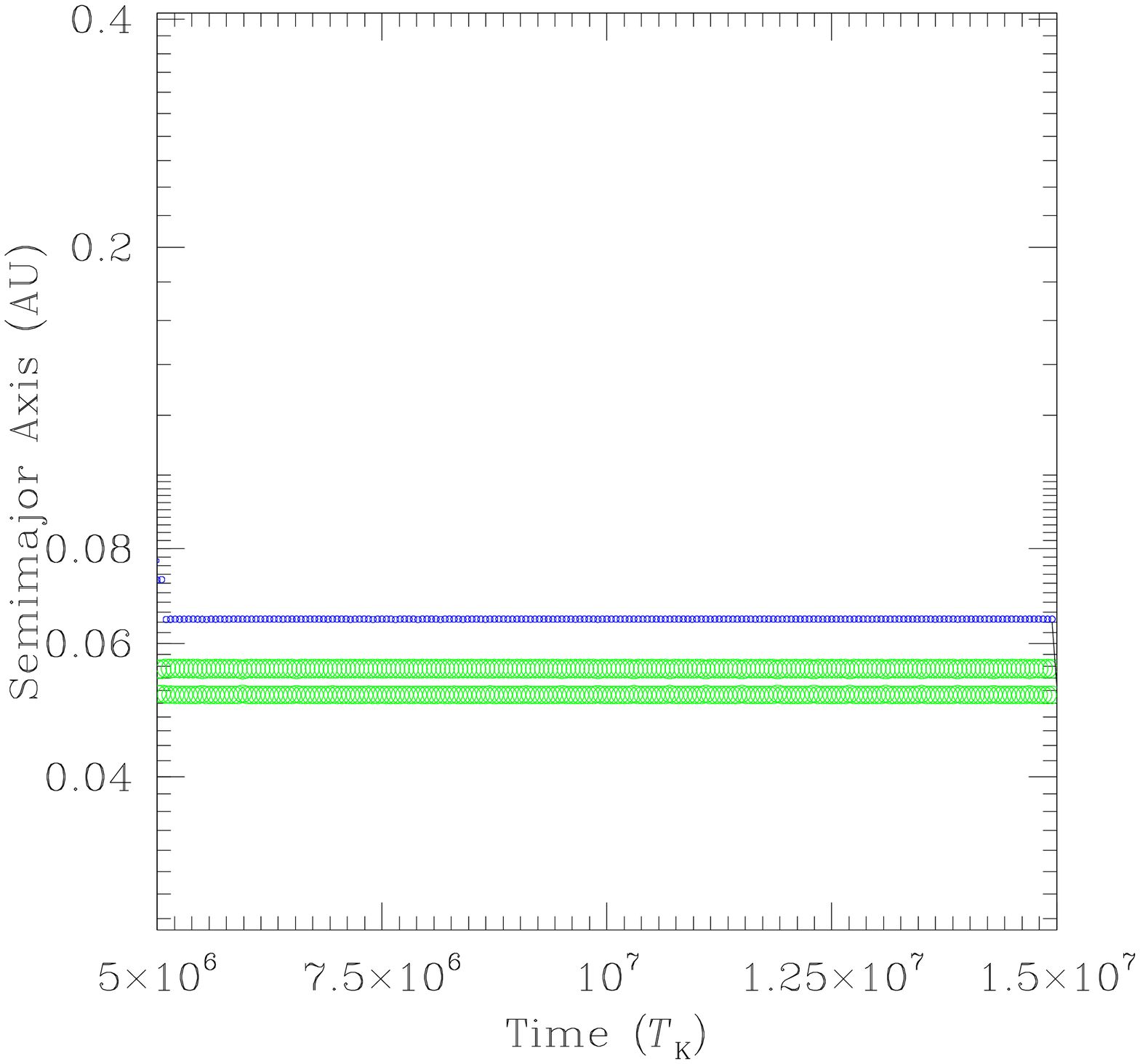}
\caption{Orbital evolution of runC1. \textit{Left}: The result of \textit{N}-body simulation 
with the effect of disk gas. \textit{Right}: Successive simulation of runC1 after the removal of disk gas.
The effect of reversed torque due to inverse pressure gradient near the disk edge is included.}
\label{fig:t_a_model53}
\end{center}
\end{figure}

\begin{figure}
\begin{center}
\epsscale{1.0}
\plottwo{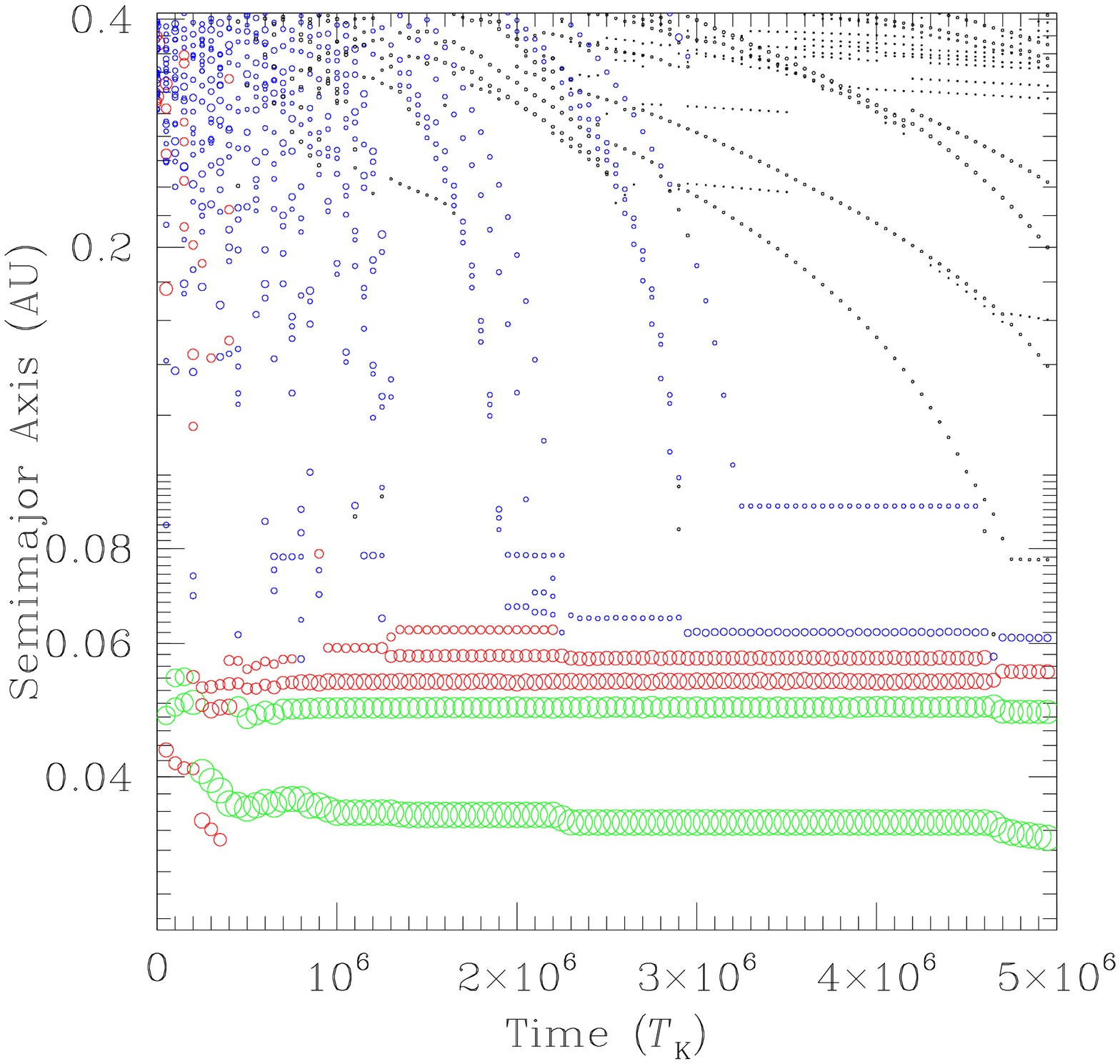}{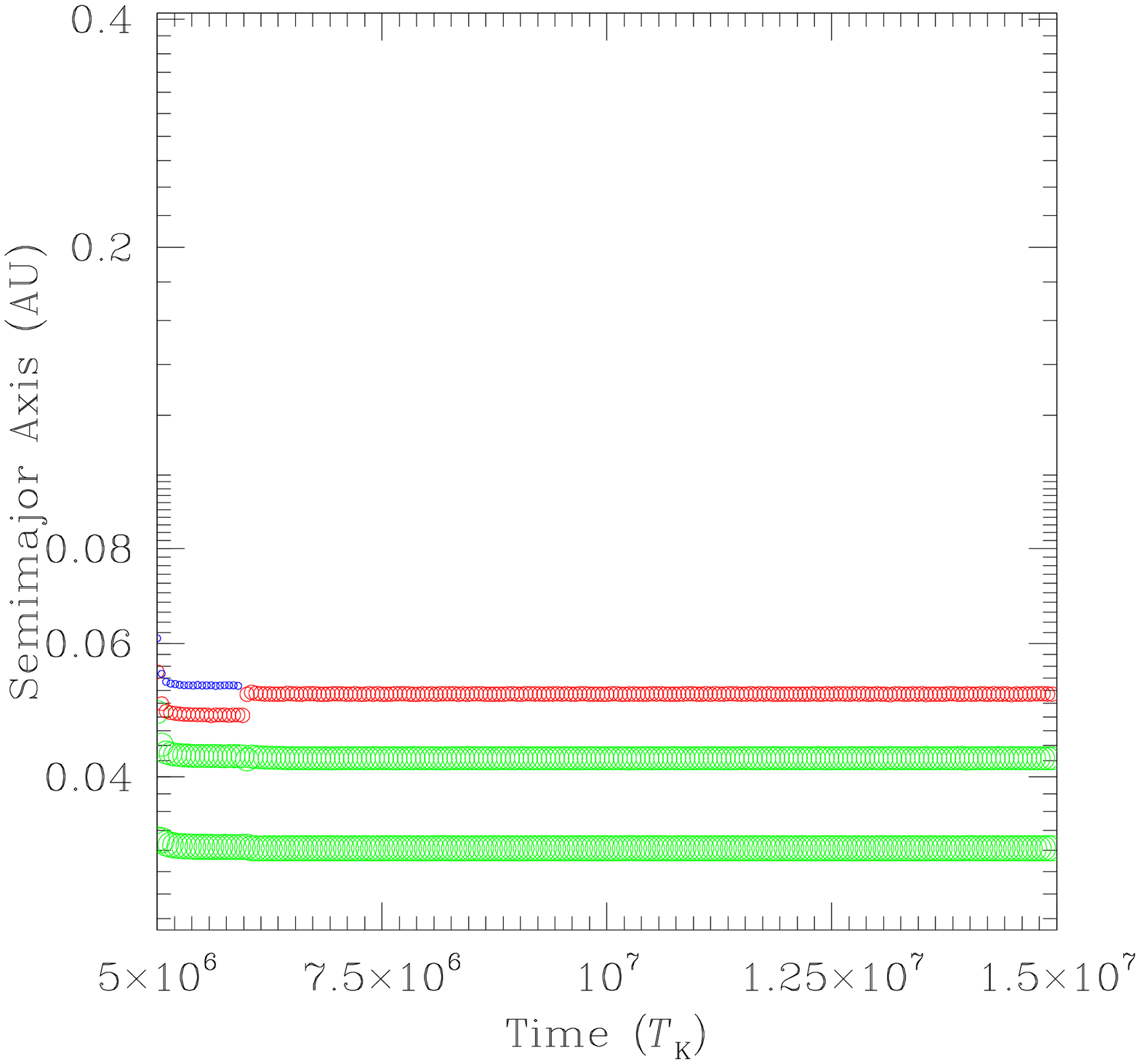}
\caption{The same as Fig.~\ref{fig:t_a_model53} for runD1. 
The enhanced ice condensation factor ($\eta_{\rm ice}=14$) is adopted.}
\label{fig:t_a_model37}
\end{center}
\end{figure}

\begin{figure}
\begin{center}
\epsscale{1.0}
\plottwo{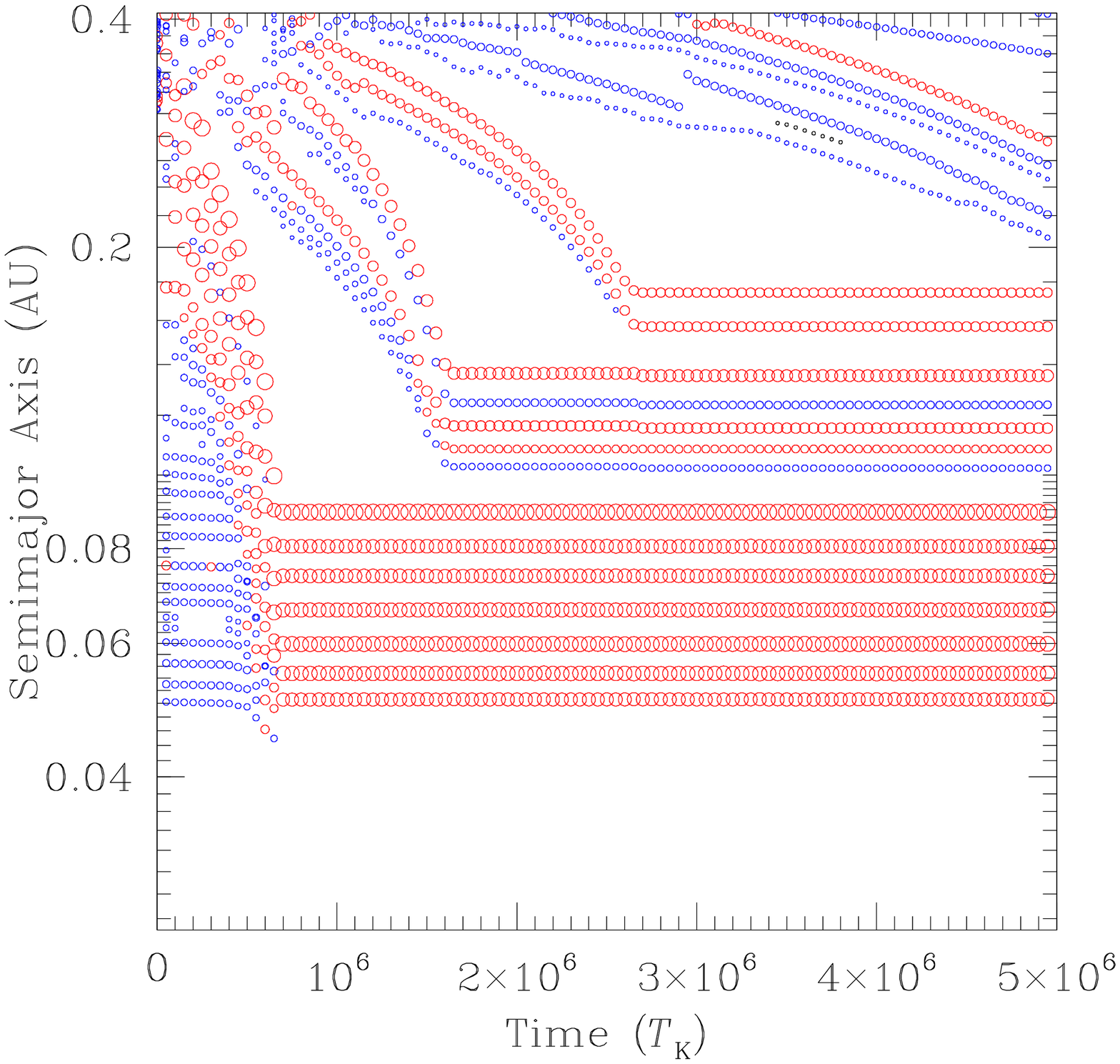}{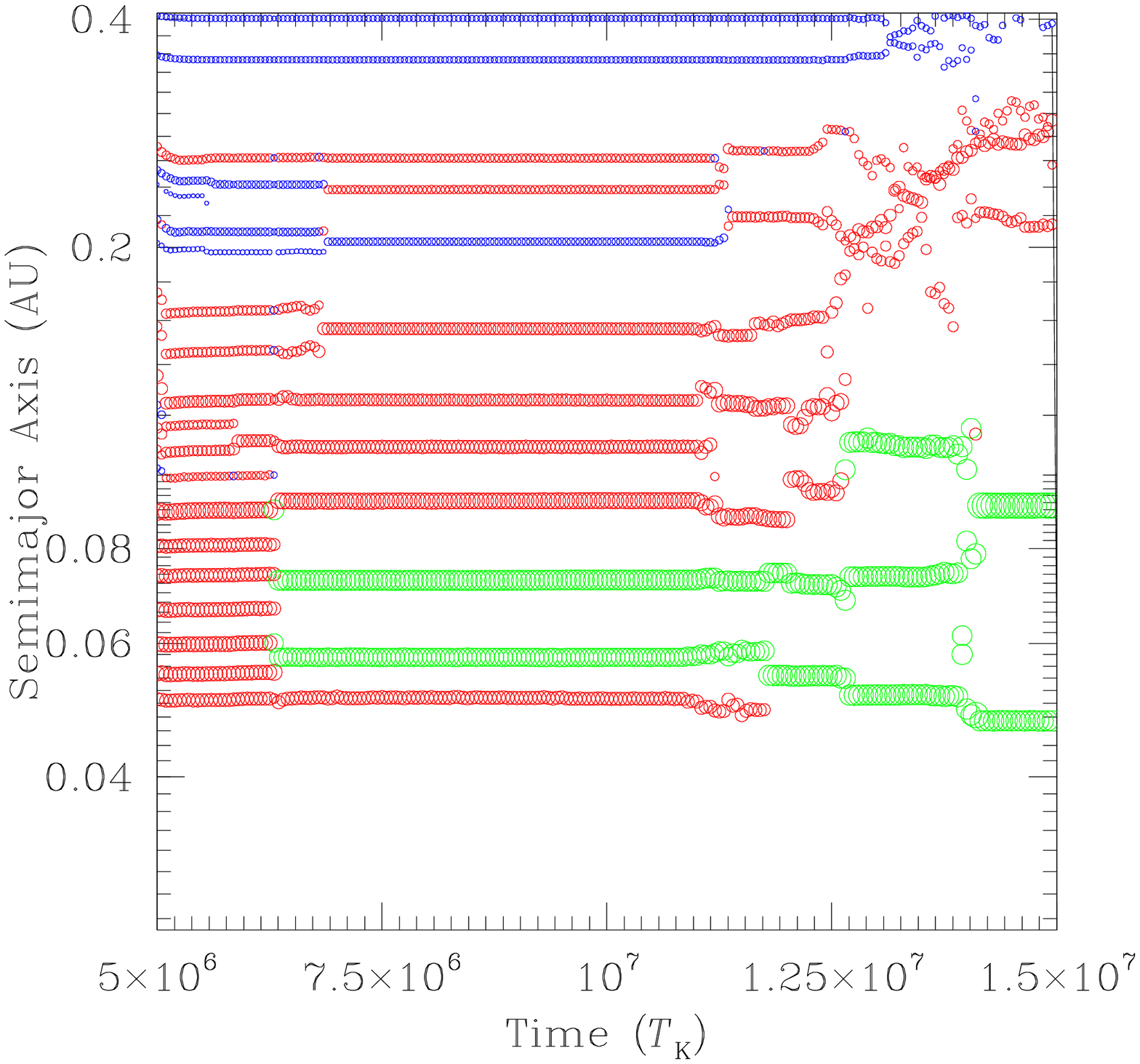}
\caption{The same as Fig.~\ref{fig:t_a_model37} for runE1.
The enhanced ice condensation factor ($\eta_{\rm ice}=14$) is adopted, 
while type-I migration is not included.}
\label{fig:t_a_model38}
\end{center}
\end{figure}

\begin{figure}
\begin{center}
\epsscale{1.0}
\plottwo{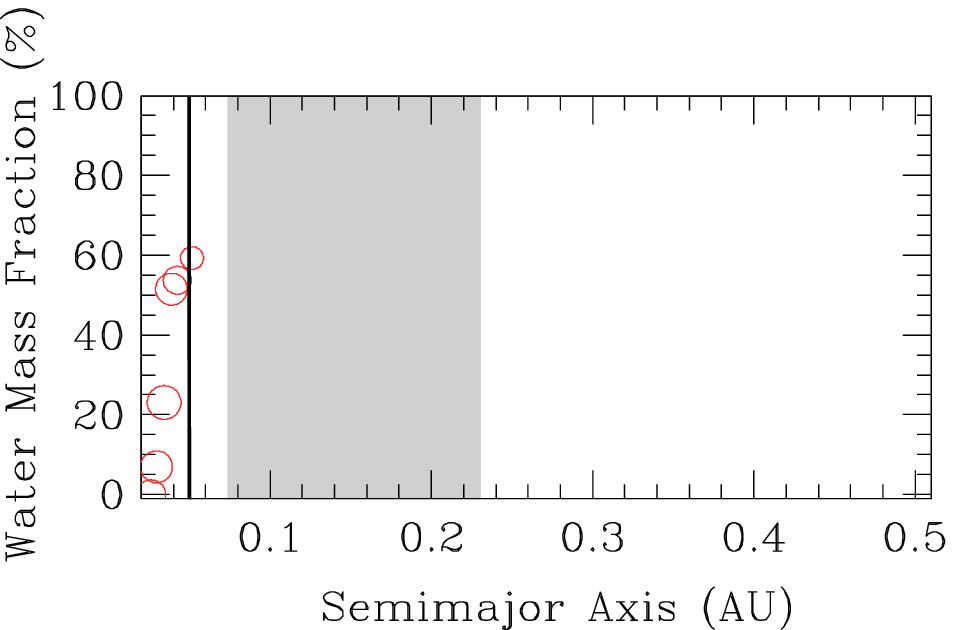}{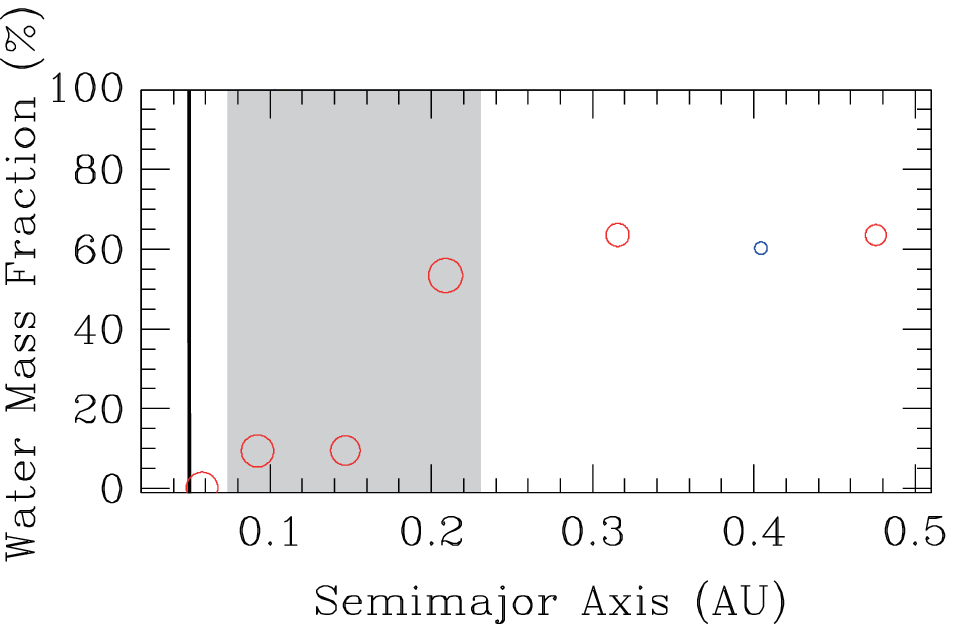}
\caption{Water mass fraction of the final planets for runA1 (\textit{left}) and runB1 (\textit{right}) 
after gas dissipation. 
The shaded regions express the HZ around a $0.01~L_{\odot}$ star.
Vertical lines represent the location of the disk inner edge (0.05~AU).
}
\label{fig:comp}
\end{center}
\end{figure}

\clearpage

\begin{deluxetable}{c|ccc|cc|ccccc}
\tabletypesize{\small}
\tablecaption{Initial Conditions and Final Results for Each Run\label{tbl:initial}}
\startdata
\hline
\hline
run & type-I &reversal&$\eta_{\rm ice}$& $N$ & $M_{\rm max}$ ($M_{\oplus}$) & $N_{\rm dis}$ & $M_{\rm max,dis}$ ($M_{\oplus}$) &$\overline{\Delta a} (r_{\rm H})$&$\overline e$& MMR \\ \hline
runA1&Yes&No&3&6&0.63&6&0.63&$8.3$&$0.0053$&Yes\\ 
runA2&Yes&No&3&5&0.79&5&0.79&$9.5$&$0.0064$&Yes\\ 
runA3&Yes&No&3&5&0.75&4&1.0&$11$&$0.019$&Yes\\ 
runA4&Yes&No&3&4&1.1&3&1.1&$9.3$&$0.0036$&Yes\\ 
runB1&No&-&3&45&0.21&11&0.65&$18$&$0.086$&No\\ 
runB2&No&-&3&38&0.24&17&0.41&$16$&$0.031$&No\\ 
runB3&No&-&3&40&0.15&14&0.53&$22$&$0.056$&No\\ 
runB4&No&-&3&36&0.21&14&0.42&$19$&$0.048$&No\\ 
runC1&Yes&Yes&3&5&1.4&3&1.4&$5.6$&$0.0084$&Yes\\ 
runC2&Yes&Yes&3&4&1.6&4&1.6&$13$&$0.011$&Yes\\ 
runD1 & Yes &No&14& 3 & 3.4 &3&3.4&$8.2$&$0.0083$&Yes\\ 
runD2 & Yes &No&14& 3 & 3.4 &3&5.7&$11$&$0.010$&Yes\\ 
runD3 & Yes &No&14& 4 & 3.6 &4&3.6&$10$&$0.024$&Yes\\ 
runD4 & Yes &No&14& 3 & 4.1 &3&4.1&$9.3$&$0.0034$&Yes\\ 
runE1 & No &-&14& 13 & 0.88 &12&3.4&$18$&$0.19$&No\\ 
runE2 & No &-&14& 17 & 1.0 &15&1.4&$16$&$0.073$&No\\ 
runE3 & No &-&14& 12 & 0.94 &17&2.3&$16$&$0.020$&No\\ 
runE4 & No &-&14& 18 & 0.83 &12&2.7&$18$&$0.13$&No\\ 
\enddata
\tablecomments{
Calculation conditions and final results. 
The second to forth columns are calculation conditions. 
The fifth and sixth columns are results of \textit{N}-body simulations before the removal of gas.
$N$ is the number of final planets, the masses of which are $\geq 0.01 M_{\oplus}$, and  
$M_{\rm max}$ is the mass of the largest planet.
The seventh to eleventh columns are results of the long term evolution after the gas removal.
$N_{\rm dis}$ is the number of final planets, the masses of which are $\geq 0.01 M_{\oplus}$, and
$M_{\rm max,dis}$ is the mass of the largest planet, 
$\overline{\Delta a}$ is the average orbital separations of final planets and
$\overline e$ is the average orbital eccentricity.
The final column expresses commensurate relationships: 
``Yes'' means that there exists at least one commensurate relationship between the final planets, while 
``No'' means that there exists no commensurate relationship.}
\end{deluxetable}


\begin{thebibliography}{}
\bibitem[Aarseth et al.(1993)]{aarseth_etal93}
Aarseth, S. J., Lin, D. N. C., \& Palmer, P. L. 1993,
\apj, 403, 351
\bibitem[Adachi et al.(1976)]{adachi_etal76}
Adachi, I., Hayashi, C., \& Nakazawa, K. 1976,
Prog. Theor. Phys., 56, 1756
\bibitem[Artymowicz(1993)]{artymowicz93} 
Artymowicz, P. 1993,
\apj, 419, 166
\bibitem[Balbus \& Hawley(1991)]{balbus91} 
Balbus, S. A., \& Hawley, J. F. 1991,
\apj, 376, 214
\bibitem[Baruteau \& Masset(2008)]{baruteau08}
Baruteau, C., \& Masset, F. 2008,
in IAU Symp. No. 249, Exoplanets: Detection, Formation and Dynamics, ed. Y.-S. Sun, S. Ferraz-Mello, \& J.-L. Zhou (Cambridge: Chambridge Univ. Press), 393

\bibitem[Beaulieu et al.(2006)]{beaulieu_etal06}
Beaulieu, J.-P., 
Bennett, D.P., et al. 2006, 
\nat, 439, 437
\bibitem[Boss(2006)]{boss06}
Boss, A. P. 2006,
\apj, 664, L79
\bibitem[Butler et al.(2004)]{butler_etal04}
Butler, R. P., Vogt, S. S., Marcy, G. W., Fischer, D. A., Wright, J. T., Henry, G. W., Laughlin, G., \& Lissauer, J. J. 2004, 
\apj, 617, 580
\bibitem[Butler et al.(2006)]{butler_etal06}
Butler, R. P., et al. 2006,
\apj, 646, 505
\bibitem[Chambers et al.(1996)]{chambers_etal96} 
Chambers, J. E., Wetherill, G. W., \& Boss, A. P. 1996, 
\icarus, 119, 261
\bibitem[Ciesla \& Cuzzi(2006)]{ciesla06}
Ciesla, F. J., \& Cuzzi, J. N. 2006
\icarus, 181, 178
\bibitem[Deming et al.(2007)]{deming_etal07} 
Deming, D., Harrington, J., Laughlin, G., Seager, S., Navarro, S. B., Bowman, W. C., \& Horning, K. 2007, 
\apj, 667, L199
\bibitem[Endl et al.(2006)]{endl_etal06}
Endl, M., Cochran, W. D., K\"{u}rster, M., Paulson, D. B., Wittenmyer, R. A., MacQueen, P. J., \& Tull, R. G. 2006, 
\apj, 649, 436
\bibitem[Fromang et al.(2005)]{fromang_etal05}
Fromang, S., Terquem, C., \& Nelson, R. P. 2005
\mnras, 363, 943
\bibitem[Gillon et al.(2007)]{gillon_etal07}
Gillon, M., et al. 2007,
\aap, 472, L13
\bibitem[Gladman(1993)]{gladman93}
Gladman, B. 1993, 
\icarus, 106, 247
\bibitem[Goldreich \& Tremaine(1980)]{goldreich80}
Goldreich, P., \& Tremaine, S. 1980,
\apj, 241, 425
\bibitem[Greenberg et al.(1978)]{greenberg_etal78}
Greenberg, R., Wacker, J. F., Hartmann, W. K., \& Chapman, C. R. 1978,
\icarus, 35, 1
\bibitem[Habets \& Heintze(1981)]{habets81}
Habets, G. M. H. J., \& Heintze, J. R. W. 1981, 
Astron. Astrophys. Suppl. Ser., 46, 193
\bibitem[Hayashi(1981)]{hayashi81}
Hayashi, C. 1981
Prog. Theor. Phys. Suppl., 70, 35
\bibitem[Hayashi et al.(1985)]{hayashi_etal85}
Hayashi, C., Nakazawa, K., \& Nakagawa, Y. 1985,
in Protostars and Planets II, ed. D. C. Blank, \& M. S. Mathews, (Tucson: Univ. of Arizona Press), 1100
\bibitem[Ida(1990)]{ida90}
Ida, S. 1990,
\icarus, 88, 129
\bibitem[Ida \& Makino(1992)]{ida92}
Ida, S., \& Makino, J. 1992,
\icarus, 96, 107
\bibitem[Ida \& Lin(2004)]{ida04}
Ida, S., \& Lin, D. N. C. 2004,
\apj, 604, 388
\bibitem[Ida \& Lin(2005)]{ida05}
Ida, S., \& Lin, D. N. C. 2005,
\apj, 626, 1045
\bibitem[Ida \& Lin(2008a)]{ida08a}
Ida, S., \& Lin, D. N. C. 2008a,
\apj, 673, 487
\bibitem[Ida \& Lin(2008b)]{ida08b}
Ida, S., \& Lin, D. N. C. 2008b,
\apj, 685, 584
\bibitem[Ikoma \& Genda(2006)]{ikoma06}
Ikoma, M., \& Genda, H. 2006,
\apj, 648, 696
\bibitem[Iwasaki et al.(2002)]{iwasaki02}
Iwasaki, K., Emori, H., Nakazawa, K., \& Tanaka, H. 2002,
Publ. Astron. Soc. Jpn., 54, 471
\bibitem[Johnson et al.(2007)]{johnson_etal07}
Johnson, J. A., Butler, R. P., Marcy, G. W., Fischer, D. A., Vogt, S. S., Wright, J. T. \& Peek, K. M. G. 2007,
\apj, 670, 833
\bibitem[Kasting et al.(1993)]{kasting_etal93}
Kasting, J. F., Whitmire, D. P., \& Reynolds, R. T. 1993, 
\icarus, 101, 108
\bibitem[Kidder(1995)]{kidder95}
Kidder, L. E. 1995,
Phys. Rev. D., 52, 821
\bibitem[Kokubo \& Ida(1995)]{kokubo95}
Kokubo, E., \& Ida, S. 1995,
\icarus, 114, 247
\bibitem[Kokubo \& Ida(1996)]{kokubo96}
Kokubo, E., \& Ida, S. 1996,
\icarus, 123, 180
\bibitem[Kokubo \& Ida(1998)]{kokubo98}
Kokubo, E., \& Ida, S. 1998,
\icarus, 131, 171
\bibitem[Kokubo \& Ida(2000)]{kokubo00}
Kokubo, E., \& Ida, S. 2000,
\icarus, 143, 15
\bibitem[Kokubo \& Ida(2002)]{kokubo02}
Kokubo, E., \& Ida, S. 2002,
\apj, 581, 666
\bibitem[Kominami \& Ida(2002)]{kominami02}
Kominami, J., \& Ida, S. 2002,
\icarus, 157, 43
\bibitem[Kretke \& Lin(2007)]{kretke07}
Kretke, K., \& Lin, D. N. C. 2007,
\apj, 664, L55
\bibitem[Kretke et al.(2009)]{kretke_etal09}
Kretke, K., Lin, D. N. C., Garaud, P., \& Turner, N. J. 2009,
\apj, 690, 407
\bibitem[Laughlin et al.(2004a)]{laughlin_etal04a}
Laughlin, G., Steinacker, A., \& Adams, F. C. 2004, 
\apj, 608, 489
\bibitem[Laughlin et al.(2004b)]{laughlin_etal04b}
Laughlin, G., Bodenheimer, P., \& Adams, F. C. 2004, 
\apj, 612, L73
\bibitem[Li et al.(2005)]{li_etal05}
Li, H., Li, S., Koller, J., Wendroff, B. B., Liska, R., Orban, C. M., Liang, E. P. T., \& Lin, D. N. C. 2005,
\apj, 624, 1003
\bibitem[Lissauer(2007)]{lissauer07}
Lissauer, J. J. 2007,
\apj, 660, L149
\bibitem[Makino(1991)]{makino91}
Makino, J. 1991,
\apj, 369, 200
\bibitem[Makino \& Aarseth(1992)]{makino92}
Makino, J., \& Aarseth, S. J. 1992, 
Publ. Astron. Soc. Jpn. 44, 141
\bibitem[Mardling \& Lin(2002)]{mardling02}
Mardling R. A., \& Lin, D. N. C. 2002,
\apj, 573, 829
\bibitem[Masset et al.(2006)]{masset_etal06}
Masset, F. S., D'Angelo, G., \& Kley, W. 2006,
\apj, 652, 730
\bibitem[Mayor et al.(2009)]{mayor_etal08}
Mayor, M., et al. 2009,
\aap, 493, 639
\bibitem[McNeil et al.(2005)]{McNeil_etal05}
McNeil, D., Duncan, M., \& Levison, H. F. 2005,
\aj, 130, 2884
\bibitem[Morbidelli et al.(2000)]{morbidelli_etal00}
Morbidelli, A., Chambers, J., Lunine, J. I., Petit, J. M., Robert, F., Valsecchi, G. B., \& Cyr, K. E. 2000,
Meteoritics \& Planetary Science, 35, 1309
\bibitem[Nagasawa et al.(2005)]{nagasawa_etal05}
Nagasawa, M., Lin, D. N. C., \& Thommes, E. 2005,
\apj, 635, 578
\bibitem[Nelson \& Papaloizou(2004)]{nelson04} 
Nelson, R. P., \& Papaloizou, J. C. B. 2005, 
\mnras, 350, 849
\bibitem[Nelson(2005)]{nelson05} 
Nelson, R. P. 2005, 
\aap, 443, 1067
\bibitem[Ogihara et al.(2007)]{ogihara_etal07}
Ogihara, M., Ida, S., \& Morbidelli, A. 2007, 
\icarus, 188, 522
\bibitem[Pollack et al.(1994)]{pollack_etal94}
Pollack, J. B., Hollenbach, D., Beckwith, S., Simonelli, D. P., Roush, T., \& Fong, W. 1994,
\apj, 421, 615
\bibitem[Raymond et al.(2007)]{raymond_etal07}
Raymond, S. N., Scalo, J., \& Meadows, V. S. 2007, 
\apj, 669, 606
\bibitem[Robert(2001)]{robert01}
Robert, F. 2001,
Science, 293, 1056
\bibitem[Scalo et al.(2007)]{scalo_etal07}
Scalo, J., et al. 2007,
Astrobiology, 7, 85
\bibitem[Selsis et al.(2007)]{selsis_etal07}
Selsis, F., Kasting, J.F., Levrard, B., Paillet, J., Ribas, I., \& Delfosse, X. 2007,
\aap, 476, 1373
\bibitem[Stevenson \& Lunine(1988)]{stevenson88}
Stevenson, D. J., \& Lunine, J. I. 1988
\icarus, 75, 146
\bibitem[Tanaka et al.(2002)]{tanaka_etal02} 
Tanaka, H., Takeuchi, T., \& Ward, W. R. 2002,
\apj, 565, 1257
\bibitem[Tanaka \& Ward(2004)]{tanaka04} 
Tanaka, H., \& Ward, W. R. 2004,
\apj, 602, 388
\bibitem[Tarter et al.(2007)]{tarter_etal07}
Tarter, J. C., et al. 2007,
Astrobiology, 7, 30
\bibitem[Terquem \& Papaloizou(2007)]{terquem07}
Terquem, C., \& Papaloizou, J. C. B. 2007,
\apj, 654, 1110
\bibitem[Udry et al.(2007)]{udry_etal07}
Udry, S., et al. 2007,
\aap, 469, L43
\bibitem[Ward(1986)]{ward86} 
Ward, W. R. 1986,
\icarus, 67, 164
\bibitem[Ward(1993)]{ward93} 
Ward, W. R. 1993,
\icarus, 106, 274
\bibitem[Ward(1997)]{ward97} 
Ward, W. R. 1997,
\apj, 482, L211
\bibitem[Wetherill \& Stewart(1989)]{wetherill89}
Wetherill, G. W., \& Stewart, G. R. 1989,
\icarus, 77, 330
\bibitem[Zhou et al.(2005)]{zhou05}
Zhou, J.-L., Aarseth, S. J., Lin, D. N. C., \& Nagasawa, M. 2005,
\apj, 631, L85

\end{thebibliography}
\end{document}